\documentclass[useAMS,usenatbib]{mn2e}
\usepackage{graphicx}
\usepackage{latexsym,epsfig}
\usepackage{longtable}
\usepackage{pdfpages}
\usepackage{epstopdf}

\newcommand{\mnras}{MNRAS}
\newcommand{\aj}{AJ}
\newcommand{\pasp}{PASP}
\newcommand{\apj}{ApJ}
\newcommand{\aap}{A\&A}
\newcommand{\aapr}{A\&ARv}
\newcommand{\araa}{ARA\&A}
\newcommand{\apjs}{ApJS}

\title[SRd variables]{Chemical compositions of  semi-regular variable red giants }

\author[Pozhath \& Lambert ]{Ramya Pozhath\thanks{e-mail:
ramyap09@gmail.com} and David L. Lambert\thanks{e-mail:dll@astro.as.utexas.edu}\\
The W.J. McDonald Observatory \& The Department of Astronomy, The University of Texas at Austin, Austin, TX 78712, USA  }

\date{Accepted XXX. Received YYY; in original form ZZZ}

\pubyear{2023}
\begin{document}
\label{firstpage}
\pagerange{\pageref{firstpage}--\pageref{lastpage}}
\maketitle

\begin{abstract}
A sample of warm  low level semi-regular variables chosen from the {\it General Catalogue of Variable Stars}  is studied for their chemical compositions by analysing  high resolution optical spectra. The abundance ratios  from Na/Fe to Eu/Fe displayed by these and previously analysed semi-regular variables are quite  similar  to  ratios displayed  by normal red giants across the Galactic thin and thick disks  and  halo populations in the solar neighbourhood suggesting  from this perspective that the variables may be among the more photometrically active red giants.
\end{abstract}


\section{Introduction}

The {\it General Catalogue of Variable Stars} defines the SRd class to be composed of `Semiregular variable giants and supergiants of F, G or K spectral types, sometimes with emission lines in their spectra. Amplitudes of light variations are in the range from 0.1 to 0.4 mag and the range of periods  is from 
50 to 1100 days' \citep{Samus2017}. Our present selection of  SRd and SRd: variables observed for spectroscopic analysis and discussion is drawn from the 2022 June version of the {\it GCVS}.\footnote{ Caution! The electronic  {\it GCVS}  does not recognize lower case letters and, for example, identifies a SRd as a SRD variable. We prefer the designation SRd but occasionally employ SRD believing the reader will appreciate their equivalence.}   Assignment to the SRd: class exploits the traditional astronomical use of a colon to indicate a variable very likely similar to a SRd variable but presumably of less certain photometric variability with respect to amplitude and/or period.  The {\it GCVS} catalogue consulted by us lists  177 SRd and 100 SRd: variables.
Among the listed variables from  the {\it GCVS}  are supergiants -- evolved massive stars -- such as R Pup and  $\rho$ Cas. Another subsample are post-AGB low mass stars such as 89 Her and UU Her with their  fascinating spectroscopic signatures.  But the majority of SRd and SRd: variables  appear to be  lower-mass giants associated with  the Red Giant Branch (RGB) including  cooler and  more luminous giants extending into the M spectral types and thus populating the upper reaches of  the RGB to its tip where He is ignited in the star's interior and the star quickly assumes the identity of a red clump (RC) giant before evolving to the AGB.  Giants on the RC and the AGB also fall within the SRd and SRd: domain. Our emphasis here is on  RGB, RC and early-AGB giants, more specifically excluding stars with spectra in which TiO bands are prominent, i.e., our sample excludes M-type giants.

Identification of a red giant as a variable  is dependent on its  photometric variability and, of course, also on a  capability to detect  this variability.  Early additions to the SRd and SRd: lists of red giants often exhibited  a visual magnitude amplitude of 1 magnitude or more and a spectral type variation, say typically G2 to K2 with excursion into the M giants. Variability linked to radial pulsations fed by astero-seismology is of  low amplitude in K giants but increases in M giants.  Other and speculative  triggers for variability may exist such as a close stellar companion or near-merging planet, a magnetic field and the stimulus resulting from He-core ignition at the RGB tip which may persist for a time in an AGB giant.

In this paper, we explore the similarity between a collection of SRd (SRd:) variables and non-variable red giants with respect to  chemical composition as anticipated from their stage of evolution  and Galactic   kinematics.  Our variable giants include stars of  differing metallicities. The Galactic characteristics of the variable and non-variable giants range from thin disk to thick disk to halo.  Our high-resolution optical spectra were collected over the years as gaps arose in our other observing programmes undertaken at the W.J. McDonald Observatory.  The principal goal was to search for  anomalies -- chemical or kinematical --  relative to   red giants not (yet?) labelled as variable. 

\section{Our sample of semi-variable red giants}

Our  observed  SRd and SRd: variables drawn from the {\it GCVS} electronic catalogue  were sifted to reject the few stars that were not  red giants and those variables with TiO bands in their spectrum.  The analysed sample  (Table \ref{warm_srd_table}) is dominated by   K giants  with typical line widths indicating a pairing of a low projected rotational velocity with typical turbulence (i.e., an Arcturus-looking optical spectrum). The  Table \ref{warm_srd_table} entries include the  name and variable type from the {\it GCVS}, the HIP number \citep{Perry1997} or an alternative identifier, the range in brightness with the minimum and maximum magnitudes which includes {\it Hipparcos} magnitudes Hp or other entries such as the visual, photovisual or Johnson $V$ magnitude range and the photographic magnitude $p$, taken from the {\it GCVS}. The Table's final column provides either the period of the variability from the {\it Hipparcos} entry or that catalogue's characterization of the variability: C = constant, M= possible microvariable with an amplitude of less than 0.03 mag. and   U = unsolved (uncertain?) variable which does not fall in the other categories, which also includes irregular or semi-regular variables, and possibly variables with amplitude not exceeding  0.03 mag. In two cases, the period of variability is taken from the {\it GCVS}.  Of the entries in Table \ref{warm_srd_table}, ten were listed as {\it GCVS} variables following {\it Hipparcos}.

Approximately half of the   variable red giants observed at McDonald  provided spectra with TiO bands present to differing degrees. In some cases, the TiO bands  may range from absent to prominent as a star executes its variation -- see, for example, the case of W LMi  (\cite{Giridhar2000}, Fig. 2). TiO bands are an additional challenge to the quantitative spectroscopist. \cite{Giridhar2000} analysed a  spectrum of AG Aur in which TiO bands were `strong'  to find  a near-normal mix of elements but  some elements were inaccessible thanks to line blending and some were unusually uncertain,   as judged by  their [M/Fe] index. Here, we postpone analysis of those variables exhibiting TiO bands.

 Three  additional variables observed at McDonald -- V897 Her, V395 Cyg and OX Ser --are not listed  in Table \ref{warm_srd_table}. Each is unusually broad-lined  suggestive of a high projected rotational velocity. We have not proceeded with a detailed abundance analysis of the trio in light of the inevitably lower accuracy of [M/Fe] indices for these broad-lined stars.

In later interpretation   of compositions of semi-regular red giants, we extend the sample to include variables previously reported in the literature as having   spectroscopically--determined chemical compositions. The chosen analyses are considered to be generally similar in precision to ours. This selection  extends the size of the analyzed sample of SRd and SRd: red giants.

\begin{table}
\caption{\label{warm_srd_table}Our SRd/SRd: sample.} 
\begin{tabular}{lp{0.6 cm}lccp{0.7 cm}} 
 \hline
 \\
   Name   & Type &   Identifier           & MinMag-MaxMag & Period   \\
     &           &                        &   (GCVS)    & (days)         \\
 \\ \hline 
 \\
   V354 And    & SRd:                 & HIP 2651            & 8.78-8.65(Hp) & U    \\
   AC Aqr	&  SRd             & 	TIC 248564916     & 10.50-10.00(V)& 68.0$^a$   \\
  LS Aqr	&  SRd:                & HIP 114426         & 8.61-8.33(V) &  U  \\
  AY Ari	&  SRd:                & HIP 12600            & 6.84-6.82(V) &  C$^b$           \\
  EU Dra	&  SRd:                & HIP 74280           & 8.76-8.56(V) &  U   \\  
  VW Dra	&   SRd:               & HIP 84496          & 7.00-6.00(V) &                \\
  HP Eri	& SRd:                 & HIP 21648           & 8.53-8.44(Hp) & U    \\
  V894 Her    &  SRd                 & HIP 80302          & 8.32-8.17(Hp)&  U                \\
  V991 Her   &  SRd:                & HIP 90417            &  9.98-9.33(Hp)   & 40.48 \\
  BN Lyn	&  SRd:                & HIP 41075           & 4.27-4.21(V) &  M               \\
  CD Psc	&  SRd:                & HIP 1938         & 9.99-9.81(Hp) &  37.21      \\
  EZ UMa	&    SRd:              & HIP 46247         & 6.28-6.23(V) &  U     \\
  KR Vir	&  SRd:                & HIP 61899          & 9.65-9.39(Hp) & U   \\
  RX Vir	&   SRd:               & TIC 96199381         & 9.10-8.70(p) &  200.0$^a$   \\

\\ \hline

\end{tabular} \\
\footnotesize{a : The photometric period is from the {\it GCVS}.} \\
\footnotesize{b : \cite{Henry2000}'s photometry suggests a 120 day period and an amplitude of 0.02 mag.}
\end{table}

\section{The {\it Robert G. Tull} Spectra}

High-resolution optical spectra of semi-regular variables were observed with the Tull coud\'{e} echelle spectrograph at the 2.7 meter Harlan J. Smith Telescope of the W.J. McDonald Observatory \citep{Tull95}  in breaks in  observing runs dedicated  to other projects.  Variables were selected from the {\it GCVS} lists of SRd and SRd: variables \citep{Samus2017}.\footnote{GCVS database, Version 2022 June.}

Spectra covered the interval 3800 to 10000 \AA\  with echelle orders incompletely recorded on the TK3 CCD longward of about 5800 \AA. Wavelength calibration 
was provided by an exposure of a Th-Ar hollow-cathode lamp. Exposures of a  flat-field lamp provided sensitivity and pixel-to-pixel calibration. The two-dimensional 
CCD exposures were reduced to a one-dimensional relative flux versus wavelength spectrum using the {\it Image Reduction and Analysis Facility (IRAF)}.\footnote{ {\it IRAF} 
is distributed by the National Optical Astronomy Observatory, which is operated by the Association of Universities for Research in Astronomy (AURA) under cooperative
agreement with the National Science Foundation.}  The resolving power was 60,000 and the S/N ratio at the centre of an order contributing  
stellar absorption lines was typically 100 or greater.

\section{Abundance analysis of  the  variables}

A  LTE  absorption  line analysis was conducted employing model atmospheres and equivalent width analysis. All selected absorption lines were chosen from our `solar' line list (Table 2 --see below). Atmospheric parameters -- effective surface temperature (T$_{\rm{eff}}$), surface gravity (log\,$g$),  microturbulent velocity ($\xi_{t}$) and metallicity ([M/H])  - were determined from a grid of model atmospheres.   Kurucz model atmospheres with no convective overshoot were used.  The code MOOG in its 2017 version was employed \citep{Sneden1973} in the procedure.

A selection of Fe\,{\sc i} and Fe\,{\sc ii} lines were selected covering a range in excitation potential  and equivalent widths (3m\AA -- 175m\AA).  The effective surface temperature T$_{\rm{eff}}$ was obtained by the method of excitation balance for Fe\,{\sc i} lines by zeroing  the slope of the Fe abundance versus Lower Excitation Potential (LEP). The microturbulent velocity $\xi_{t}$ was fixed by making the Fe\,{\sc i} abundances independent of the line strength. Surface gravity log\,$g$ was estimated demanding ionization balance, i.e.,  the Fe  abundance  from neutral and ionized Fe lines be the same. Finally, individual elemental abundances were estimated  using the measured equivalent widths of the unblended spectral lines and the selected model atmosphere corresponding to the adopted stellar atmospheric parameters. For transitions with significant hyperfine splitting (Sc, V, Mn, Co, Ba and Eu), hyperfine corrections were applied using the {\it blends} driver in MOOG. The wavelengths and relative strengths of the hyperfine components were taken from Kurucz database except for Ba, for which they are  taken from \cite{McWilliam1998}. The Sun  provides a  reference spectrum for our analysis. Also, we undertook an analysis of Arcturus, a red giant  with atmospheric parameters  similar to some  of our variables. Our inferred composition of Arcturus is assessed relative to published analyses of the star.

\subsection{The Sun and Arcturus}

Our solar abundance analysis is based on  the solar spectrum provided by \cite{Hinkle2000} and the Kurucz (1998) model solar atmosphere. 
Our  line list in Table 2 was compiled from various  sources, namely \cite{Bensby2003,  Bubar2010,Giridhar1994,Hekker2007,Morel2014,Pomp2011,Ramirez2011,Ramya2019,Reddy2003}, as well as the Kurucz database\footnote{https://lweb.cfa.harvard.edu/amp/ampdata/kurucz23/sekur.html}. Hyperfine wavelength splittings were 
applied for Sc, V, Mn, Co, Ba and Eu.  
The Sun was considered using the 2017 version of the LTE spectral line analysis and spectrum synthesis code MOOG \citep{Sneden1973}. 
 Analysis of the line list  gave  the solar atmosphere parameters: temperature T$_{\rm eff} = 5777$ K and  surface gravity $\log g$ = 4.44 cm s$^{-2}$  and 
microturbulent velocity $\xi_{t}$ = 1.14 km s$^{-1}$.

 Mean elemental abundances for each atom and ion with N, the number of contributing lines, are given in Table \ref{sun_abu} which also  gives  \cite{Asplund2009} spectroscopic  estimates for  solar elemental abundances.  The mean difference $|Current - Asplund|$ is a  $0.03\pm0.06$ dex.  Our iron abundance commonly adopted as `the solar metallicity' is 7.44$\pm$0.06 from the Fe \,{\sc i} lines and 7.45$\pm$0.06  from Fe\,{\sc ii} lines  values slightly less than Asplund's Fe abundance of 7.50$\pm$0.04. Our  derived solar abundances are taken as the reference for the  abundance analysis of the variable stars.

Since our investigation is directed  at K giants,  we  sought to compare our results against those for the K giant Arcturus. In this study, we limit our  study  to elements above Na but anticipate studying lighter elements and their isotopes in a later exercise.  For the Arcturus spectrum, we adopt  
that provided by \cite{Hinkle2000}. Atmospheric parameters were derived  
from the atoms and ions well represented in our line list.   These parameters given in Table \ref{arct_model_param} are the effective temperature (T$_{\rm eff}$),  the surface gravity ($g$, usually expressed as $\log g$), the microturbulence ($\xi$ assumed to be isotropic and depth independent) and the composition ( [M/H] with relative elemental abundances taken to be solar). Our  results and two reference results are given in Table \ref{arct_model_param}. There may be some overlap in methods and stellar and atomic data between the three analyses but, nonetheless, the close agreement between the analyses is pleasing.

Elemental abundances from our analysis   are provided 
in Table \ref{arct_abu} where we list $\log \epsilon$(M) with its scatter $\sigma$ and  the number of measured lines $N$,  the abundance ratio [M/Fe] relative to the Sun.  
In the  final columns,  we give [M/Fe] as determined by the  reference studies in Table \ref{arct_model_param}. Our abundances are  consistent with  previous studies: For example, the differences in [M/Fe] between our analysis and that by  \cite{Ramirez2011} run from $+0.08$ to $-0.12$ for a mean of $0.00$. A detailed  comparison of the studies would be required to identify precise  reasons for the small differences. All of the reported studies are based on an LTE analysis.

\subsection{K giant semi-regular variables}

 Stars in Table \ref{warm_srd_table} were observed and analysed.  Atmospheric parameters  for individual stars are given in Table \ref{atm_warmSRD}. Elemental abundances as [M/Fe] with their standard deviation and number of contributing lines  are provided 
 in Table \ref{abu_warmSRD}. Iron abundances from Fe\,{\sc i} and Fe\,{\sc ii} lines are both given relative to the solar Fe abundance. Also provided in the Tables is the assigned Galactic population -- halo, thick or thin disk (see below). Uncertainties in the atmospheric parameters are estimated to be $\Delta$T$_{\rm{eff}}$ = $\pm$50 K, $\Delta$log\,$g$ = $\pm$0.2 dex, $\Delta$ $\xi_{t}$ = $\pm$0.2 km s$^{-1}$, and $\Delta$[M/H] = $\pm$0.1 dex (see \cite{Ramya2012} and \cite{Ramya2016}). The uncertainty in the abundances due to any parameter is estimated by measuring the amount by which the mean abundances vary responding to a change in the respective parameter by an amount equal to the
uncertainty in the parameter, keeping the other parameters constant. Following the method, we estimated
the resultant uncertainties in the abundances, $\sigma$(T$_{\rm eff}$), $\sigma$(log $g$), $\sigma$($\xi_{t}$) and $\sigma$([M/H]) due to the uncertainties in the atmospheric parameters $\Delta$T$_{\rm{eff}}$, $\Delta$log\,$g$, $\Delta$ $\xi_{t}$ and $\Delta$[M/H] respectively. The results obtained by applying this procedure to one of our sample stars AY Ari is given in Table \ref{sensi_AYAri}. Assuming that the uncertainties in the parameters are unrelated
and independent, the net uncertainty in the abundance ratio was calculated as a quadratic sum, which is given in the last column as $\sigma$(model). It is seen that the scatter for a given  [M/Fe] for AY Ari  in  Table \ref{abu_warmSRD} is  similar to that from  Table \ref{sensi_AYAri}.

\onecolumn
\begin{longtable}{lccccc}
  \caption{Our linelist} \\
  \hline
   Ion &  Wavelength    & LEP    &  log $gf$  &  W$_\lambda\odot$  &  log $\varepsilon_\odot$ \\
       &   (\AA)        &  (eV)    &         &  (m\AA)   &  (dex) \\\hline \\

Na I&  6154.23   & 2.100 &  -1.550 &  36.6  &  6.26  \\
&  6160.75   & 2.100 &  -1.250 &  56.5  &  6.28     \\

Mg I&  5711.09  & 4.340 &  -1.730 &  104.1  &  7.56   \\
&  6318.72   & 5.110 &  -1.950 &  44.5  &  7.57        \\
&  6319.24   & 5.110 &  -2.320 &  27.4  &  7.63        \\
&  7657.61   & 5.110 &  -1.280 &  98.5  &  7.61        \\

Al I&  6696.02   & 3.140 &  -1.480 &  36.9  &  6.38   \\
&  6698.67   & 3.140 &  -1.780 &  20.8  &  6.33        \\
&  7835.31   & 4.020 &  -0.690 &  41.1  &  6.40        \\
&  7836.13   & 4.020 &  -0.450 &  55.0  &  6.36      \\

Si I&  5690.42   & 4.930 &  -1.770 &  48.1  &  7.45  \\
&  5701.10   & 4.930 &  -1.950 &  37.9  &  7.45        \\
&  5772.15   & 5.080 &  -1.650 &  52.3  &  7.54       \\
&  6142.49   & 5.620 &  -1.540 &  33.3  &  7.57        \\
&  6145.02   & 5.610 &  -1.479 &  37.6  &  7.58        \\

Ca I&  5260.39   & 2.520 &  -1.720 &  32.1  &  6.25  \\
&  5867.56   & 2.930 &  -1.570 &  22.6  &  6.25        \\
&  6166.44   & 2.520 &  -1.140 &  69.1  &  6.32        \\
&  6169.04   & 2.520 &  -0.800 &  90.3  &  6.34        \\
&  6169.56   & 2.530 &  -0.480 &  108.7  &  6.30      \\
&  6455.60   & 2.520 &  -1.340 &  55.9  &  6.27      \\
&  6471.66   & 2.530 &  -0.690 &  90.6  &  6.22     \\
&  6499.65   & 2.520 &  -0.820 &  84.7  &  6.25     \\

Sc II&  5357.20   & 1.510 &  -2.110 &  4.8    &  3.13  \\
&  5552.23   & 1.460 &  -2.280 &  4.6    &  3.22      \\
&  5684.21   & 1.510 &  -1.070 &  37.1  &  3.17        \\
&  6245.64   & 1.510 &  -1.040 &  35.2   &  3.08      \\
&  6300.75   & 1.510 &  -1.950 &  8.2    &  3.18      \\
&  6320.84   & 1.500 &  -1.920 &  8.9    &  3.18      \\

Ti I&  5295.77   & 1.070 &  -1.580 &  13.2  &  4.89  \\
&  5490.15   & 1.460 &  -0.880 &  22.0  &  4.84        \\
&  5702.66   & 2.290 &  -0.590 &  7.3  &  4.78     \\
&  5716.44   & 2.300 &  -0.720 &  5.9  &  4.82     \\
&  6092.79   & 1.890 &  -1.320 &  4.1  &  4.84     \\
&  6303.75  & 1.440 &  -1.510 &  8.8  &  4.93     \\
&  6312.23   & 1.460 &  -1.500 &  8.1  &  4.90    \\
&  6599.10   & 0.900 &  -2.030 &  9.3  &  4.92     \\
&  7357.73   & 1.440 &  -1.070 &  22.2  &  4.92     \\
Ti II&  4583.41   & 1.170 &  -2.870 &  33.0  &  5.02   \\
&  4708.66   & 1.240 &  -2.370 &  53.3  &  5.02       \\
&  5336.78   & 1.580 &  -1.630 &  71.7  &  4.96       \\
&  5418.77   & 1.580 &  -2.110 &  48.5  &  4.94      \\

V I&  6039.73   & 1.060 &  -0.650 &  12.3   &  3.87  \\
&  6081.44   & 1.050 &  -0.580 &  13.0   &  3.80       \\
&  6090.21   & 1.080 &  -0.060 &  32.3   &  3.83       \\
&  6119.53   & 1.060 &  -0.320 &  21.0   &  3.82       \\
&  6135.36   & 1.050 &  -0.750 &  10.2   &  3.85       \\
&  6274.65   & 0.270 &  -1.670 &   6.9   &  3.80       \\

Cr I&  5287.20   & 3.440 &  -0.890 &  11.3  &  5.61  \\
&  5300.74   & 0.980 &  -2.080 &  59.3  &  5.56       \\
&  5304.18   & 3.460 &  -0.680 &  16.0  &  5.59     \\
&  5628.62   & 3.420 &  -0.760 &  14.7  &  5.58       \\
&  5781.16   & 3.010 &  -1.000 &  16.7  &  5.49       \\
&  6882.48   & 3.440 &  -0.380 &  32.4  &  5.65        \\
&  6883.00   & 3.440 &  -0.420 &  30.5  &  5.64    \\

Mn I&  4671.69   & 2.890 &  -1.660 &  14.8   &  5.42  \\
&  4739.11   & 2.940 &  -0.600 &  60.7   &  5.36      \\
&  5004.89   & 2.920 &  -1.640 &  14.0   &  5.40     \\

Fe I&  5054.64   & 3.640 &  -2.140 &  39.1  &  7.55   \\
&  5088.16   & 4.150 &  -1.680 &  36.9  &  7.52      \\
&  5198.72   & 2.220 &  -2.140 &  94.9  &  7.36       \\
&  5253.02   & 2.280 &  -3.940 &  18.7  &  7.53       \\
&  5285.13   & 4.430 &  -1.640 &  27.9  &  7.54      \\
&  5294.55   & 3.640 &  -2.760 &  14.5  &  7.53       \\
&  5295.31   & 4.420 &  -1.590 &  28.5  &  7.49       \\
&  5322.00   & 2.280 &  -2.840 &  60.1  &  7.35     \\
&  5358.12   & 3.300 &  -3.162 &  9.50  &  7.39         \\
&  5379.57   & 3.690 &  -1.510 &  60.1  &  7.38      \\
&  5386.33   & 4.150 &  -1.670 &  31.4  &  7.38        \\
&  5441.34   & 4.310 &  -1.630 &  30.4  &  7.47      \\
&  5487.76   & 4.140 &  -0.710 &  86.8  &  7.49      \\
&  5543.94   & 4.220 &  -1.140 &  61.3  &  7.52        \\
&  5565.71   & 4.610 &  -0.230 &  84.4  &  7.35        \\
&  5569.63   & 3.420 &  -0.540 &  140.4  &  7.43       \\
&  5572.85   & 3.400 &  -0.310 &  172.0  &  7.45        \\
&  5576.10   & 3.430 &  -1.000 &  111.3  &  7.56       \\
&  5586.77   & 3.400 &  -0.210 &  194.9  &  7.51       \\
&  5522.45   & 4.209 &  -1.450 &  42.6  &  7.45        \\
&  5546.51   & 4.372 &  -1.210 &  50.5  &  7.51        \\
&  5560.21   & 4.430 &  -1.090 &  51.0  &  7.46       \\
&  5577.02   & 5.030 &  -1.550 &  11.5  &  7.50       \\
&  5661.35   & 4.280 &  -1.756 &  21.8  &  7.35      \\
&  5662.51   & 4.180 &  -0.570 &  89.9  &  7.43       \\
&  5638.26   & 4.220 &  -0.770 &  75.8  &  7.43        \\
&  5679.02   & 4.652 &  -0.750 &  58.2  &  7.45      \\
&  5618.63   & 4.209 &  -1.280 &  49.7  &  7.42       \\
&  5652.33   & 4.261 &  -1.850 &  26.1  &  7.54        \\
&  5701.55   & 2.560 &  -2.220 &  83.6  &  7.47        \\
&  5705.47   & 4.302 &  -1.500 &  37.7  &  7.48       \\
&  5731.76   & 4.260 &  -1.200 &  56.5  &  7.52        \\
&  5732.28   & 4.990 &  -1.560 &  14.1  &  7.58        \\
&  5753.13   & 4.260 &  -0.760 &  76.2  &  7.44       \\
&  5775.08   & 4.221 &  -1.141 &  58.4  &  7.46       \\
&  5778.45   & 2.588 &  -3.440 &  22.2  &  7.40      \\
&  5793.91   & 4.220 &  -1.619 &  33.2  &  7.42       \\
&  5849.69   & 3.695 &  -2.930 &  6.6  &  7.34       \\
&  5855.08   & 4.608 &  -1.478 &  21.0  &  7.35       \\
&  5856.10   & 4.294 &  -1.558 &  32.5  &  7.41        \\
&  5858.79   & 4.220 &  -2.180 &  12.5  &  7.40      \\
&  5859.60   & 4.550 &  -0.620 &  69.3  &  7.43       \\
&  5862.37   & 4.550 &  -0.420 &  82.4  &  7.46       \\
&  5902.47   & 4.593 &  -1.750 &  13.7  &  7.38      \\
&  5905.67   & 4.650 &  -0.690 &  56.8  &  7.36       \\
&  5916.25   & 2.453 &  -2.990 &  54.0  &  7.49       \\
&  5927.79   & 4.650 &  -0.990 &  41.5  &  7.36        \\
&  5929.68   & 4.550 &  -1.310 &  39.5  &  7.55        \\
&  5934.65   & 3.930 &  -1.170 &  73.2  &  7.50      \\
&  5984.83   & 4.710 &  -0.300 &  80.5  &  7.44     \\
&  6003.01   & 3.880 &  -1.060 &  81.7  &  7.49      \\
&  6008.54   & 3.870 &  -1.010 &  85.5  &  7.50       \\
&  6015.25   & 2.220 &  -4.680 &  5.10  &  7.52      \\
&  6024.07   & 4.530 &  -0.120 &  103.8  &  7.46         \\
&  6027.05   & 4.076 &  -1.300 &  63.2  &  7.56        \\
&  6056.00   & 4.730 &  -0.400 &  70.6  &  7.39         \\
&  6065.49   & 2.610 &  -1.530 &  114.4  &  7.34        \\
&  6078.49   & 4.796 &  -0.388 &  73.6 &  7.47        \\
&  6079.01   & 4.650 &  -1.020 &  44.9  &  7.46         \\
&  6093.64   & 4.607 &  -1.410 &  30.2  &  7.51       \\
&  6096.66   & 3.984 &  -1.810 &  36.9  &  7.46         \\
&  6098.24   & 4.558 &  -1.800 &  15.8  &  7.46      \\
&  6082.72   & 2.220 &  -3.570 &  34.0  &  7.43        \\
&  6120.25   & 0.920 &  -5.950 &  5.2  &  7.49        \\
&  6127.91   & 4.140 &  -1.400 &  47.6  &  7.41        \\
&  6137.00   & 2.200 &  -2.950 &  65.2  &  7.42         \\
&  6137.72   & 2.590 &  -1.400 &  123.0  &  7.31         \\
&  6151.62   & 2.176 &  -3.300 &  48.7  &  7.40        \\
&  6159.38   & 4.610 &  -1.830 &  12.6  &  7.42       \\
&  6165.36   & 4.143 &  -1.460 &  43.9  &  7.40        \\
&  6173.34   & 2.220 &  -2.880 &  67.0  &  7.40        \\
&  6180.21   & 2.730 &  -2.650 &  53.3  &  7.40      \\
&  6187.99   & 3.940 &  -1.620 &  46.1  &  7.40         \\
&  6213.44   & 2.220 &  -2.480 &  81.7  &  7.31       \\
&  6219.29   & 2.200 &  -2.430 &  87.9  &  7.36      \\
&  6240.65   & 2.223 &  -3.287 &  47.8  &  7.41      \\
&  6252.55   & 2.404 &  -1.891 &  118.3  &  7.55        \\
&  6265.14   & 2.180 &  -2.550 &  85.2  &  7.40       \\
&  6270.23   & 2.858 &  -2.540 &  51.6  &  7.37         \\
&  6271.28   & 3.330 &  -2.703 &  23.8  &  7.41         \\
&  6322.70   & 2.590 &  -2.430 &  74.5  &  7.47       \\
&  6335.35   & 2.200 &  -2.350 &  96.0  &  7.42        \\
&  6336.82   & 3.686 &  -0.856 &  100.9  &  7.42         \\
&  6344.15   & 2.430 &  -2.920 &  61.4  &  7.52         \\
&  6380.75   & 4.190 &  -1.380 &  50.5  &  7.49         \\
&  6392.54   & 2.280 &  -4.030 &  17.8  &  7.53        \\
&  6411.65   & 3.640 &  -0.720 &  120.6  &  7.52         \\
&  6419.96   & 4.710 &  -0.240 &  81.3  &  7.38        \\
&  6436.41   & 4.186 &  -2.360 &  10.0  &  7.42        \\
&  6475.63   & 2.560 &  -2.940 &  54.4  &  7.52  \\
&  6481.88   & 2.280 &  -2.980 &  63.0  &  7.46       \\
&  6498.94   & 0.960 &  -4.690 &  45.4  &  7.47         \\
&  6518.40   & 2.830 &  -2.460 &  56.2  &  7.35         \\
&  6574.23   & 0.990 &  -5.000 &  27.7  &  7.45        \\
&  6581.21   & 1.480 &  -4.680 &  20.8  &  7.45        \\
&  6591.33   & 4.593 &  -1.950 &  10.5  &  7.42         \\
&  6593.89   & 2.430 &  -2.420 &  83.8  &  7.46         \\
&  6608.03   & 2.280 &  -4.030 &  17.2  &  7.50        \\
&  6609.11   & 2.560 &  -2.690 &  64.0  &  7.46       \\
&  6677.98   & 2.690 &  -1.470 &  121.0  &  7.39         \\
&  6699.14   & 4.590 &  -2.100 &  7.9  &  7.42      \\
&  6703.57   & 2.759 &  -3.023 &  36.1  &  7.42        \\
&  6705.10   & 4.607 &  -0.980 &  45.4  &  7.37         \\
&  6710.32   & 1.490 &  -4.870 &  15.1  &  7.47        \\
&  6713.75   & 4.795 &  -1.400 &  20.8  &  7.42         \\
&  6725.36   & 4.103 &  -2.167 &  17.0  &  7.41         \\
&  6726.67   & 4.607 &  -1.030 &  46.0  &  7.43        \\
&  6733.15   & 4.638 &  -1.400 &  25.9  &  7.40        \\
&  6750.16   & 2.420 &  -2.620 &  72.7  &  7.41        \\
&  6793.26   & 4.076 &  -2.326 &  12.4  &  7.37         \\
&  6810.26   & 4.607 &  -0.986 &  48.6  &  7.43        \\
&  6828.59   & 4.640 &  -0.820 &  54.6  &  7.41     \\
&  6837.01   & 4.590 &  -1.687 &  17.6  &  7.41     \\
&  6842.69   & 4.640 &  -1.220 &  39.1  &  7.51        \\
&  6843.66   & 4.550 &  -0.830 &  59.5  &  7.43        \\
&  6857.25   & 4.076 &  -2.038 &  22.3  &  7.40         \\
&  6911.51   & 2.420 &  -4.040 &  13.3  &  7.50        \\
&  6976.93   & 4.580 &  -1.850 &  16.5 &  7.53        \\
&  6971.94   & 3.020 &  -3.340 &  12.6  &  7.36        \\
&  6999.88   & 4.100 &  -1.460 &  53.8  &  7.51        \\
&  7022.95   & 4.190 &  -1.150 &  63.6  &  7.48      \\
&  7132.99   & 4.080 &  -1.650 &  42.1  &  7.45        \\
&  7802.51   & 5.080 &  -1.310 &  15.4  &  7.39       \\
&  7807.92   & 4.990 &  -0.509 &  58.8  &  7.45         \\
Fe II&  4491.40  & 2.856 &  -2.700 &  77.5  &  7.51  \\
&  4508.29   & 2.856 &  -2.440 &  84.6  &  7.41    \\
&  4576.33   & 2.844 &  -2.950 &  63.4  &  7.39        \\
&  4620.52   & 2.830 &  -3.310 &  50.9  &  7.42        \\
&  4670.18   & 2.583 &  -4.090 &  30.8  &  7.48                                     \\
&  5197.57   & 3.230 &  -2.220 &  79.3  &  7.36                                     \\
&  5234.62   & 3.221 &  -2.180 &  83.3  &  7.39                                     \\
&  5256.92   & 2.890 &  -4.060 &  20.7  &  7.45                                     \\
&  5264.80   & 3.230 &  -3.130 &  47.9  &  7.53                                     \\
&  5325.60   & 3.220 &  -3.220 &  41.4  &  7.46                                     \\
&  5414.07   & 3.221 &  -3.580 &  27.5  &  7.47                                     \\
&  5425.26   & 3.200 &  -3.220 &  41.3  &  7.43                                     \\
&  5534.84   & 3.250 &  -2.750 &  58.0  &  7.40                                     \\
&  5991.38   & 3.150 &  -3.540 &  30.7  &  7.44                                     \\
&  6084.10   & 3.200 &  -3.780 &  20.6  &  7.45                                     \\
&  6113.32   & 3.220 &  -4.110 &  11.1  &  7.45                                     \\
&  6149.25   & 3.889 &  -2.630 &  35.7  &  7.36                                     \\
&  6247.54   & 3.890 &  -2.430 &  52.8  &  7.56                                     \\
&  6369.46   & 2.891 &  -4.110 &  19.4  &  7.45                                     \\
&  6416.93   & 3.890 &  -2.640 &  38.9  &  7.44                                     \\
&  6432.68   & 2.890 &  -3.690 &  40.7  &  7.57                                     \\
&  6456.39   & 3.903 &  -2.065 &  62.3  &  7.41                                     \\
&  6516.05   & 2.890 &  -3.310 &  52.3  &  7.45                                     \\
&  7515.83   & 3.900 &  -3.460 &  13.2  &  7.49                                     \\
&  7711.72   & 3.904 &  -2.555 &  46.1  &  7.49                                     \\

Co I&  5280.63  & 3.630 &  -0.030 &  20.3   &  4.82  \\
&  5352.04  & 3.580 &   0.060 &  25.1   &  4.79                                     \\
&  5647.23  & 2.280 &  -1.560 &  13.9   &  4.83                                     \\
&  6455.00  & 3.630 &  -0.250 &  14.8   &  4.82                                     \\

Ni I&  5088.96  & 3.678 &  -1.240 &  28.2  &  6.16  \\
&  5094.42  & 3.833 &  -1.074 &  30.4  &  6.19                                     \\
&  5115.40  & 3.834 &  -0.281 &  75.2  &  6.30                                     \\
&  6111.08  & 4.088 &  -0.808 &  33.6  &  6.20                                     \\
&  6130.14  & 4.266 &  -0.938 &  21.6  &  6.21                                     \\
&  6175.37  & 4.089 &  -0.550 &  47.7  &  6.23                                     \\
&  6176.80  & 4.090 &  -0.260 &  62.0  &  6.22                                     \\
&  6177.25  & 1.826 &  -3.508 &  14.1  &  6.17                                     \\
&  6772.32  & 3.658 &  -0.972 &  48.0  &  6.22                                     \\
&  7797.59  & 3.900 &  -0.348 &  75.3  &  6.29                                     \\
&  7826.77  & 3.700 &  -1.840 &  12.5  &  6.20                                     \\

Zn I&  4810.54  & 4.080 &  -0.170 &  71.6  &  4.48  \\
&  6362.35  & 5.790 &  0.140 &  21.2  &  4.52        \\

Y II&  5200.39  & 0.990 &  -0.570 &  36.3  &  2.10   \\
&  5289.82  & 1.030 &  -1.850 &  3.8  &  2.16                                      \\
&  5402.76  & 1.840 &  -0.620 &  11.6  &  2.24                                   \\
&  4883.67  & 1.080 &  0.070 &  56.5  &  2.07                                      \\
&  4982.11  & 1.030 &  -1.290 &  13.4  &  2.23                                     \\

Zr I&  4739.47  & 0.650 &  0.230 &  6.7  &  2.51    \\
&  6143.21  & 0.070 &  -1.100 &  2.3  &  2.68        \\

Ba II&  5853.68  & 0.604 &  -1.000 &  60.3  &  2.17  \\
&  6141.73  & 0.704 &  -0.032 & 109.1  &  2.31      \\

La II&  6390.48  & 0.320 &  -1.410 &  2.7  &  1.09   \\
&  4662.49  & 0.000 &  -1.240 &  7.0  &  1.14                                     \\
&  5303.54  & 0.320 &  -1.350 &  3.9  &  1.25                                     \\
&  6774.24  & 0.130 &  -1.820 &  2.6  &  1.27                                     \\

Nd II&  5319.81  & 0.550 &  -0.140 &  10.9  &  1.39   \\
&  5092.79  & 0.380 &  -0.610 &  5.5  &  1.37                                     \\
&  4989.92  & 0.630 &  -0.310 &  6.9  &  1.43                                     \\

Eu II&  6645.13  & 1.380 &  0.204 &  5.4    &  0.53  \\

\\  \hline
\end{longtable}

\twocolumn

\begin{table}
\caption{\label{sun_abu}The solar abundances.}
\begin{tabular}{p{0.6 cm}lllc}
\hline \\
Ion & N &    Mean $\pm \sigma$  & Asp09$^{a}$  &  Mean-Asp09  \\
    &   &    (dex)              &    (dex)     &  (dex)       \\ \hline

\hline \\

Na\,{\sc i}  & 2 &  6.27 $\pm$0.02 &  6.24 $\pm$ 0.04  &  +0.03  \\
Mg\,{\sc i} & 4 &  7.59 $\pm$ 0.04  &  7.60 $\pm$ 0.04  & -0.01  \\
Al\,{\sc i} & 4 &  6.37 $\pm$ 0.03    &  6.45 $\pm$ 0.03  & -0.08  \\
Si\,{\sc i} & 5 &  7.52$\pm$  0.06    &  7.51 $\pm$ 0.03  & +0.01 \\
Ca\,{\sc i} & 8 &  6.28$\pm$ 0.04 &  6.34 $\pm$ 0.04  & -0.06  \\
Sc\,{\sc ii} & 6 &  3.16 $\pm$  0.05   &  3.15 $\pm$ 0.04  & +0.01  \\
Ti\,{\sc i} & 9 &  4.87 $\pm$ 0.05  &  4.95 $\pm$ 0.05  & -0.08  \\
Ti\,{\sc ii} & 4 &   4.98 $\pm$ 0.04  &               &  +0.03  \\
V\,{\sc i} & 6 &   3.83 $\pm$ 0.03 &  3.93 $\pm$ 0.08  &  -0.10  \\
Cr\,{\sc i} & 7 &  5.59$\pm$ 0.05    &  5.64 $\pm$ 0.04  & -0.05  \\
Mn\,{\sc i} & 3 &  5.39$\pm$  0.03   &  5.43 $\pm$ 0.04  & -0.04  \\
Fe\,{\sc i}  & 124 &  7.44 $\pm$ 0.06   &  7.50 $\pm$ 0.04  & -0.06 \\
Fe\,{\sc ii} & 25 &   7.45 $\pm$ 0.06  &             &  -0.05 \\
Co\,{\sc i} & 4 &   4.81 $\pm$0.02   &  4.99 $\pm$ 0.07  & -0.18 \\
Ni\,{\sc i} & 11 &   6.22 $\pm$0.05 &  6.22 $\pm$ 0.04  & 0.00 \\
Zn\,{\sc i} & 2 &  4.50 $\pm$0.03  &  4.56 $\pm$ 0.05  & -0.06 \\
Y\,{\sc ii} & 5 &   2.16$\pm$ 0.07   &  2.21 $\pm$ 0.05  & -0.05  \\
Zr\,{\sc i}  & 2 &  2.59$\pm$0.12  &  2.58 $\pm$ 0.04  & +0.01  \\
Ba\,{\sc ii} & 2 &  2.24 $\pm$ 0.10   &  2.18 $\pm$ 0.09  & +0.06 \\
La\,{\sc ii} & 4 & 1.19 $\pm$ 0.09    &  1.10 $\pm$ 0.04  & +0.09  \\
Nd\,{\sc ii} & 3 &  1.40$\pm$  0.03  &  1.42 $\pm$ 0.04  & -0.02  \\
Eu\,{\sc ii} & 1 &  0.53         &  0.52 $\pm$ 0.04  & +0.01 \\

\\  \hline
% \end{supertabular} \\

\end{tabular} \\
\footnotesize{a : \cite{Asplund2009}}\\
\end{table}

\begin{table}
\caption{\label{arct_model_param}Arcturus model stellar parameters}
\begin{tabular}{l c c c} 

\hline \\
 Parameter  &  Here   & R\&A11$^{a}$ & Wo09$^{b}$ \\    
\hline \\
  T$_{\rm{eff}}$ (K)     &   4285   &  4286  &  4300 \\      
  log\,$g$ (cm s$^{-2}$) &   1.50   &  1.66  &  1.60   \\    
  $\xi_{t}$ (km s$^{-1}$) &   1.72   &  1.74  &  1.50  \\    
  {}[M/H]                &  -0.59   & -0.52  &  -0.60  \\    

  \\  \hline 
\footnotesize{a : \cite{Ramirez2011}} \\ 
\footnotesize{b : \cite{Worley2009}}\\

\end{tabular} 
\end{table}

\begin{table}
\caption{\label{arct_abu}Arcturus abundances.}
\begin{tabular}{p{0.6 cm}lp{0.8cm} p{1.6 cm}c}

\hline \\
Ion  & log $\epsilon$(M) and (N) & [M/Fe]  &  [M/Fe]  & [M/Fe] \\                              %    & [M/Fe] \\ 
\cline{2-3}
     & \multicolumn{2}{c} {{Current study}}           & R\&A11$^a$ & Wo09$^b$ \\      % & Ka18$^c$ \\
\hline \\

Na\,{\sc i}    & 5.81 $\pm$ 0.01(2) &  0.13  &  	  0.11 $\pm$ 0.03    &    0.15 $\pm$  0.04 \\  %   &      0.02     \\
Mg\,{\sc i}     & 7.41 $\pm$ 0.02(4) &  0.41  &  	  0.37 $\pm$ 0.03    &    0.34 $\pm$  0.15 \\  %    &       0.48     \\
Al\,{\sc i}     & 6.15 $\pm$ 0.03(4) &  0.37 &   	  0.34 $\pm$ 0.03    &    0.25 $\pm$  0.07  \\ %   &       ...      \\
Si\,{\sc i}     & 7.24 $\pm$ 0.08(5) &  0.31 &   	  0.33 $\pm$ 0.04    &    0.24 $\pm$  0.14 \\  %   &       ...      \\
Ca\,{\sc i}     & 5.84 $\pm$ 0.03(8) &  0.15 &   	  0.11 $\pm$ 0.04    &    0.19 $\pm$  0.06  \\ %   &       ...      \\
Sc\,{\sc ii}    & 2.66 $\pm$ 0.04(6) &  0.09  &  	  0.23 $\pm$ 0.04    &    0.24 $\pm$  0.01  \\ %  &       ...       \\
Ti\,{\sc i}     & 4.51 $\pm$ 0.07(9) &  0.23 &      0.27 $\pm$ 0.05    &    0.35 $\pm$  0.12  \\   %  &       ...       \\
Ti\,{\sc ii}    & 4.54 $\pm$ 0.04(4) &  0.15  & 	  0.21 $\pm$ 0.04    &    0.33 $\pm$  0.10  \\ %  &       ...      \\
V\,{\sc i}      & 3.32 $\pm$ 0.03(6) &  0.08 &      0.20 $\pm$ 0.05    &       ...         \\      % &      ...      \\
Cr\,{\sc i}     & 5.00 $\pm$ 0.09(7) &  -0.01 &    -0.05 $\pm$ 0.04   &       ...         \\       % &       ...      \\
Mn\,{\sc i}     & 4.63 $\pm$ 0.01(3) &  -0.18  &   -0.21 $\pm$ 0.04   &       ...         \\       %  &      ...      \\
Fe\,{\sc i}     & 6.85 $\pm$ 0.08(122) &  ---  &    ---              &    ---            \\        % &       ---       \\   
Fe\,{\sc ii}    & 6.87 $\pm$ 0.07(25) &  ---  &     ---              &    ---            \\        % &       ---       \\
Co\,{\sc i}     & 4.36 $\pm$ 0.05(4) &  0.14  &     0.09 $\pm$ 0.04    &       ...         \\      %  &      ...      \\
Ni\,{\sc i}     & 5.66 $\pm$ 0.04(11) &  0.03   &   0.06 $\pm$ 0.03    &       ...         \\      %  &      ...      \\
Zn\,{\sc i}     & 4.12 $\pm$ 0.22(2) &  0.21    &   0.22 $\pm$ 0.06    &    -0.04 $\pm$ 0.09  \\   %  &      ...       \\
Y\,{\sc ii}     & 1.65 $\pm$ 0.12(4) &  0.08     &      ...        &    0.12  $\pm$ 0.11   \\      % &      -0.11     \\
Zr\,{\sc i}     & 1.80 $\pm$ 0.04(2) &  -0.21    &      ...        &    0.01  $\pm$ 0.07  \\       % &       ...      \\
Zr\,{\sc ii}    & 1.97 $\pm$ -   (1) &  -0.09    &       ...        &    0.12  $\pm$ 0.10  \\      %  &      -0.03     \\
Ba\,{\sc ii}    & 1.51 $\pm$ 0.19(2) &  -0.15   &        ...        &    -0.19 $\pm$ 0.08  \\      %  &      ....      \\
La\,{\sc ii}    & 0.47 $\pm$ 0.09(4) &  -0.14   &       ...        &    0.04  $\pm$ 0.08   \\      % &      -0.09     \\
Nd\,{\sc ii}    & 0.91 $\pm$ 0.13(3) &  0.10    &       ...        &    0.10 $\pm$  0.07  \\       %  &      -0.09     \\
Eu\,{\sc ii}    & 0.17 $\pm$ -   (1) &  0.23   &        ...        &    0.36  $\pm$ 0.04   \\      % &       0.25     \\
\\  \hline 

\end{tabular}
\footnotesize{a : \cite{Ramirez2011}} \\ 
\footnotesize{b : \cite{Worley2009}} \\
\end{table}

 \section{Galactic kinematics of the variables}

In  broad terms, the stellar population in the solar neighbourhood is considered to arise from three components: the thin disk, the thick disk and the halo. These components are definable primarily by their Galactic velocity components U,V,W. In this section, we determine the components U,V,W for our variables and, also, for similar semi-variables stars with compositions reported previously in the literature. In a subsequent section, we assess whether our and comparable previously analysed variables have compositions expected of normal giants with the same Galactic kinematics.

Each  population is represented by a three dimensional Gaussian distribution in the velocity space, with different dispersions along U, V and W. Adopting the kinematical properties for the thin disc, thick disk and halo distributions and the fraction of solar neighborhood stars belonging to each of these components from Table 1 of \cite{Reddy2006} (and references therein), we calculated the probability with which a star belongs to these three main components of the Galaxy. In a majority of cases, it was clear from the probabilities that the star's membership was assignable to  one of the three Galactic population. Among our sample of variables, the majority belong to the thin disk but the thick and halo are represented.

Our calculations of U,V,W and the probabilities of membership are based on Gaia astrometric  values \citep{Gaia2016, Gaia2023, Babusiaux2023} but otherwise follow the recipe described by \cite{Reddy2006} which was based on {\it Hipparcos} not Gaia astrometric input. Gaia's DR3 values for the stellar parallax and proper motion but the radial velocity from our {\it Tull} spectrum were used in the computation of the U,V,W velocity components with respect to the Sun following the equations given in \cite{Johnson1987}. Population assignments would not be appreciably affected were the Gaia radial velocity preferred. Additionally, a variable   surely has a somewhat variable radial velocity. A right-handed coordinate system is used where U is the radial component positive towards the Galactic center, V is the tangential component in the direction of Galactic rotation and W is the vertical component positive towards the North Galactic Pole. In general, disk kinematics imply large tangential or rotational velocities around the Galactic center (hence small with respect to Sun), while halo stars have the least rotational velocities around Galactic center and have highly eccentric orbits.

Table \ref{kine_sample} provides the kinematical parameters for our  warm  variables.   The sample is dominated by thin disk members (10 of 14) with the thick disk contributing just a single star and the halo three. Oddly, there is not a single entry with a mixed membership; presumably because our sample is too small. No attempt is made to search for kinematic structure within these Galactic populations. However, the thick disk's representative AC Aqr seems unusually metal-poor at [Fe/H] $=-2.3$ for this population.

 In considering variables drawn from the literature (see below), we use data from Gaia to compute velocities UVW and the above recipe to estimate probabilities of membership of Galactic halo, thick and thin disks.

\begin{table}
\caption{\label{atm_warmSRD}Derived  atmospheric parameters  for the sample stars grouped by Galactic population}
\begin{tabular}{lcccc}

\hline \\
 Star  &   T$_{\rm eff}$(K) & log\,$g$(cm s$^{-2}$) & $\xi_{t}$(km s$^{-1}$)  & [M/H] (dex)   \\
 %      &   (spec)    & (spec) & (spec) & (spec)   \\
 \\ \hline \\ 
 \multicolumn{5}{c}{{Thin disk stars}} \\
  V354 And  & 4360    & 1.48        & 1.76   & -0.27      \\  
   AY Ari    & 4900    & 2.98       & 1.50   & +0.31       \\   
 EU Dra    & 4880   & 2.65         & 0.92   &  -0.35     \\ 
  VW Dra   & 4670    & 2.26        & 1.58   &  -0.05      \\   
 V894 Her  & 4700    & 2.50        & 1.55   &  -0.12      \\   
  V991 Her & 4750   &  2.60         & 2.23  &  -0.15     \\ 
 BN Lyn    & 3910    & 0.98        &  1.54  &  +0.01     \\  
 CD Psc    & 4930    & 2.52        & 1.81   & -0.24       \\ 
 EZ UMa    & 4020    & 1.24        & 1.62   & -0.16      \\ 
  RX Vir   & 5860    & 3.80       & 1.31   & +0.13      \\  \\
\multicolumn{5}{c}{{Thick disk star}} \\
  AC Aqr    & 4220   &  0.07        & 2.13   &  -2.30     \\  \\
  \multicolumn{5}{c}{{Halo stars}} \\
  LS Aqr    & 4030  &  0.10         &  2.13  & -1.65       \\   
 HP Eri    & 4310   & 0.52         & 2.16   &  -1.78     \\   
 KR Vir    & 4060   & 0.05         & 2.36   &  -1.50     \\   

\\  \hline
% \end{longtable}
\end{tabular}

\end{table}

% \onecolumn

\begin{table*}
\caption{\label{abu_warmSRD}Derived  abundances for the sample stars}
\begin{center}
\begin{tabular}{lccccccc}

\hline \\
  & \multicolumn{7}{c}{{Thin disk stars}} \\
\cline{2-8}
                   &V354 And      &AY Ari       &EU Dra       & VW Dra        &V894 Her     &V991 Her      &BN Lyn \\
\hline 
\\
{}[Na\,{\sc i}/Fe] &0.47 0.06 2   &0.30 0.06 2   &0.11 0.07 2   &0.05 0.03 2   &0.24 0.04 2   &0.38 - 1  &0.11 0.07 2 \\
{}[Mg\,{\sc i}/Fe] &0.03 0.06 2   &-0.06 0.04 4  &-0.08 0.16 2   &0.12 0.07 4   &0.08 0.06 3   &0.28 0.12 2  &0.03 0.04 4 \\
{}[Al\,{\sc i}/Fe] &0.31 0.09 4   &0.07 0.05 4   &0.20 0.06 4   &0.15 0.03 4   &0.23 0.03 4   &0.36 0.09 3  &0.11 0.06 4 \\
{}[Si\,{\sc i}/Fe] &0.02 0.10 4   &0.06 0.09 5   &0.02 0.09 5   &0.11 0.07 5   &0.05 0.05 5   &0.38 0.08 2  &0.22 0.12 5 \\
{}[Ca\,{\sc i}/Fe] &0.02 0.09 6   &-0.05 0.07 8  &0.01 0.09 8   &0.02 0.04 8   &0.13 0.06 8   &0.34 0.06 5  &-0.06 0.08 8 \\
{}[Sc\,{\sc ii}/Fe]&-0.32 0.04 4  &0.02 0.05 6   &-0.31 0.04 2  &-0.04 0.05 6  &-0.08 0.06 6  &0.06 0.04 2  &-0.21 0.10 5 \\
{}[Ti\,{\sc i}/Fe] &-0.04 0.12 6  &-0.05 0.07 7  &0.03 0.1 5    &0.01 0.07 7   &0.05 0.08 6   &0.36 0.12 5  &-0.03 0.09 6 \\
{}[Ti\,{\sc ii}/Fe] &-0.24 0.09 3 &-0.17 0.1 4   &-0.39 0.09 4  &-0.08 0.06 4  &-0.21 0.07 4  &0.24 0.22 2  &-0.27 0.07 4 \\
{}[V\,{\sc i}/Fe]  &-0.04 0.07 5  &-0.06 0.05 5  &0.05 0.04 5   &-0.02 0.04 5  &0.04 0.06 5   &0.33 0.11 5  &0.14 0.08 5 \\
{}[Cr\,{\sc i}/Fe] &0.15 0.08 5   &-0.01 0.06 6  &0.12 0.12 7   &0.01 0.08 6   &0.14 0.06 6   &0.08 0.12 4  &-0.05 0.09 6 \\
{}[Mn\,{\sc i}/Fe] &-0.02 0.03 3  &-0.07 0.05 3  &-0.22 0.09 3  &-0.08 0.1 3   &-0.01 0.08 3  &0.22 -   1   &-0.34 0.04 3 \\
{}[Fe\,{\sc i}/H]&-0.27 0.11 84   &0.30 0.08 112 &-0.34 0.09 101&-0.03 0.08 103&-0.11 0.08 104&-0.13 0.13 63&-0.01 0.09 62 \\
{}[Fe\,{\sc ii}/H]&-0.27 0.11 9   &0.31 0.09 17  &-0.33 0.06 8  &-0.05 0.08 21 &-0.12 0.09 23 &-0.14 0.11 7 &0.01 0.06 6 \\
{}[Co\,{\sc i}/Fe] &-0.10 0.07 4  &-0.01 0.05 4  &-0.16 0.08 4  &0.00 0.05 4    &-0.01 0.04 4  &0.01 0.06 3  &-0.12 0.07 4 \\
{}[Ni\,{\sc i}/Fe] &-0.03 0.08 10 &0.06 0.07 11  &-0.01 0.1 10  &-0.01 0.06 11 &-0.04 0.06 11 &-0.04 0.07 8 &-0.09 0.12 7 \\
{}[Zn\,{\sc i}/Fe] &0.14 0.05 2   &0.07 0.1 2    &0.06 0.04 2   &0.10 0.08 2    &0.06 0.11 2   &-0.16 -  1  &0.35 - 1 \\
{}[Y\,{\sc ii}/Fe] &0.26 0.03 3   &0.11 0.07 5   &0.01 0.11 5   &-0.07 0.06 5  &0.27 0.14 5   &0.17 0.14 2  &0.90 0.06 2 \\
{}[Zr\,{\sc i}/Fe]  &-0.19 0.11 2 &-0.15 0.18 2  &0.04 0.01 2   &-0.24 0.09 2  &-0.02 0.16 2  &0.03 0.16 2  &0.17 0.07 2 \\
{}[Ba\,{\sc ii}/Fe] &0.26 0.18 2  &0.07 0.16 2   &0.37 0.08 2   &0.11 0.18 2   &0.28 0.23 2   &-0.02 0.12 2  &0.78 0.15 2 \\
{}[La\,{\sc ii}/Fe] &-0.15 0.08 3 &0.10 0.12 4   &0.03 0.14 4   &-0.04 0.09 4  &-0.0 0.08 3   &0.08 0.11 2   &-0.15 0.10 3 \\
{}[Nd\,{\sc ii}/Fe] &0.0 0.01 2   &0.05 0.09 3   &0.02 0.08 3   &0.09 0.08 3   &0.17 0.14 3   & ---          &0.43 0.11 3 \\
{}[Eu\,{\sc ii}/Fe] &---          &0.10 - 1      & ---          &0.17 - 1      &0.04 - 1      & ---          &0.09 - 1 \\
\hline \\
 & \multicolumn{3}{c}{{Thin disk stars}}& Thick disk star & \multicolumn{3}{c}{{Halo stars}} \\
 \cline{2-4}  \cline{6-8}
                  &CD Psc      &EZ UMa         &RX Vir       &AC Aqr        &LS Aqr        &HP Eri        &KR Vir \\
\hline 
\\
{}[Na\,{\sc i}/Fe]&0.16 0.01 2 &0.17 0.03 2   &-0.10 0.05 2   & ---         &0.04 0.02 2   &  ---         &-0.18 0.04 2 \\
{}[Mg\,{\sc i}/Fe]&0.13 0.06 2 &0.08 0.07 4   &-0.11 0.10 3    &0.32 - 1    &0.51 -    1    &0.32 0.08 2   &0.24 0.13 4 \\
{}[Al\,{\sc i}/Fe]&0.29 0.11 4 &0.29 0.05 4   &-0.09 0.02 4  &0.80 0.09 2   &0.32 0.07 3   &  ---         &0.04 - 1 \\
{}[Si\,{\sc i}/Fe]&0.10 0.08 4  &0.17 0.05 5   &-0.08 0.06 5  &0.73 0.15 2     &0.51 0.06 5   &0.39 0.06 4   &0.33 0.10 5 \\
{}[Ca\,{\sc i}/Fe]&0.11 0.08 7 &0.02 0.07 8   &0.01 0.07 7    &0.32 0.09 6  &0.28 0.04 8   &0.27 0.01 6   &0.18 0.06 7 \\
{}[Sc\,{\sc ii}/Fe]&-0.14 - 1  &-0.05 0.13 6  &-0.13 0.07 5  &-0.12 - 1     &0.04 0.04 4   &-0.13 0.04 4  &-0.11 0.07 4 \\
{}[Ti\,{\sc i}/Fe]&0.15 0.09 2 &0.00 0.08 6    &-0.07 0.03 3  &0.42 - 1     &0.24 0.07 7   &0.15 0.06 5   &0.22 0.11 6 \\
{}[Ti\,{\sc ii}/Fe]&0.10 0.11 2 &-0.11 0.08 3  &-0.06 0.05 4  &0.44 0.09 4  &0.33 0.10 4    &0.32 0.08 4   &0.30 0.10 4 \\
{}[V\,{\sc i}/Fe]&0.19 0.06 5  &0.13 0.06 5   &0.06 0.06 4   & ---          &-0.08 0.07 6   &-0.11 0.04 5  &-0.06 0.07 5 \\
{}[Cr\,{\sc i}/Fe]&0.21 0.13 5 &0.05 0.09 5   &-0.02 0.07 5  &-0.29 - 1     &-0.18 0.10 3   &-0.23 0.04 3  &-0.36 0.11 2 \\
{}[Mn\,{\sc i}/Fe]&0.11 - 1    &-0.38 0.07 3  &-0.14 0.06 2  & ---          &-0.30 - 1      & ---          &-0.34 - 1 \\
{}[Fe\,{\sc i}/H]&-0.23 0.09 82&-0.16 0.11 88 &0.14 0.08 111 &-2.30 0.09 54 &-1.66 0.09 90  &-1.77 0.07 91 &-1.50 0.09 81 \\
{}[Fe\,{\sc ii}/H]&-0.24 0.09 9&-0.18 0.11 8  &0.13 0.07 21  &-2.29 0.04 9  &-1.63 0.10 13  &-1.77 0.08 13 &-1.50 0.08 9 \\
{}[Co\,{\sc i}/Fe]&-0.06 0.11 3&-0.04 0.07 4  &-0.01 0.08 4   &              &0.06 0.09 3  &0.03 0.02 2    &-0.02 0.07 2 \\
{}[Ni\,{\sc i}/Fe]&-0.09 0.09 7&-0.04 0.10 9   &-0.06 0.07 11 &0.03 0.05 4  &-0.11 0.04 9  &-0.10 0.06 8   &-0.19 0.08 8 \\
{}[Zn\,{\sc i}/Fe]&-0.15 - 1   &0.50 - 1      &-0.07 0.02 2  &0.12 - 1      &0.28 0.06 2   &-0.01 - 1      &0.00 - 1 \\
{}[Y\,{\sc ii}/Fe]&0.29 0.12 3 &0.49 0.08 2   &0.10 0.07 5    &-0.35 0.07 2 &0.04 0.11 5   &-0.10 0.10 4   &0.11 0.06 3 \\
{}[Zr\,{\sc i}/Fe]&0.32 - 1    &-0.16 0.16 2  & ---           & ---         &0.02 0.07 2   &0.18 0.09 2   &0.01 0.01 2 \\
{}[Ba\,{\sc ii}/Fe]&0.53 0.08 2&0.54 0.17 2   &0.28 0.08 2   &-0.29 0.12 2  &-0.07 0.04 2  &0.22 0.15 2   &0.27 0.27 2 \\
{}[La\,{\sc ii}/Fe]&0.27 0.06 2&-0.06 0.12 3  &-0.01 0.12 3   &-0.17 - 1    &0.02 0.05 4   &0.17 0.06 3   &-0.05 0.07 4 \\
{}[Nd\,{\sc ii}/Fe]&0.42 - 1   &0.42 0.15 2   &0.12 0.11 3   &-0.15 0.10 3  &0.12 0.09 3   &0.27 0.08 3   &0.13 0.08 3 \\
{}[Eu\,{\sc ii}/Fe]&0.23 - 1   &0.16 - 1      &0.06 0.0 1    & ---          &0.16 - 1      &0.40 - 1      &0.36 - 1 \\
\hline

\end{tabular}
\end{center}
\end{table*}

\begin{table}
\caption{\label{sensi_AYAri}Sensitivity of the abundances ($\Delta$[M/Fe]) to AY Ari's  atmospheric parameters.}
\begin{tabular}{lccccc}
 \hline \\

Ion  & $\sigma$(T$_{\rm eff}$)  & $\sigma$(log $g$) & $\sigma$($\xi_{t}$) & $\sigma$([M/H]) & $\sigma$(model) \\
               &              &   &   &  & \\ 
\hline \\

Na\,{\sc i}  &  $\pm$0.04	  &     $\pm$0.03	 &   $\pm$0.08	&  $\pm$0.01 & 0.09 	 \\ 
Mg\,{\sc i}  &  $\pm$0.02	  &     $\pm$0.01	 &   $\pm$0.05	&  $\pm$0.01 & 0.06  		 \\ 
Al\,{\sc i}  &  $\pm$0.03	  &     $\pm$0.02	 &   $\pm$0.05	&  $\pm$0.01 & 0.06   \\ 
Si\,{\sc i}  &  $\pm$0.03	  &     $\pm$0.02	 &   $\pm$0.04	&  $\pm$0.02 & 0.06  \\ 
Si\,{\sc ii} &   $\pm$0.11  &     $\pm$0.04	 &   $\pm$0.04	&     0.0  & 0.12  		 \\ 
Ca\,{\sc i}  &  $\pm$0.04	  &     $\pm$0.03	 &   $\pm$0.11	&  $\pm$0.01 & 0.12  	 \\
Sc\,{\sc ii} &   $\pm$0.01 &     $\pm$0.08 &   $\pm$0.03	&  $\pm$0.04     & 0.09     	 \\
Ti\,{\sc i}  &  $\pm$0.07	   &       0.0	 &   $\pm$0.07	&  $\pm$0.01 & 0.10  	 \\
Ti\,{\sc ii} &   $\pm$0.01   &     $\pm$0.06 &   $\pm$0.11	&  $\pm$0.03 & 0.13  	 \\
V\,{\sc i}  &   $\pm$0.07   &     $\pm$0.0	 &   $\pm$0.07	&  $\pm$0.01 & 0.10  	 \\
Cr\,{\sc i} &   $\pm$0.04	   &     $\pm$0.01 &   $\pm$0.09	&  $\pm$0.01 & 0.10 	 \\
Mn\,{\sc i} &    $\pm$0.04  &     $\pm$0.01 &   $\pm$0.07	&  $\pm$0.01     & 0.08 	 \\
Fe\,{\sc i}  &  $\pm$0.02	   &     $\pm$0.01 &   $\pm$0.10	&  $\pm$0.01 & 0.10  	 \\
Fe\,{\sc ii} &   $\pm$0.06   &     $\pm$0.07 &   $\pm$0.09	&  $\pm$0.03 & 0.13  	 \\
Co\,{\sc i} &    $\pm$0.01  &     $\pm$0.04 &   $\pm$0.02	&  $\pm$0.02     & 0.05  \\
Ni\,{\sc i}  &    0.0	   &     $\pm$0.02 &   $\pm$0.10	&  $\pm$0.02 & 0.10 	 	 \\
Zn\,{\sc i} &   $\pm$0.03	   &     $\pm$0.04 &   $\pm$0.09	&  $\pm$0.03 & 0.11  	 \\
Y\,{\sc ii}  &  $\pm$0.01	   &     $\pm$0.07 &   $\pm$0.10	&  $\pm$0.03 & 0.13 	 \\
Zr\,{\sc i}  &  $\pm$0.08	   &     $\pm$0.0	 &   $\pm$0.05	&  $\pm$0.01 & 0.09 	 \\
Zr\,{\sc ii} &   $\pm$0.01   &     $\pm$0.08 &   $\pm$0.03	&  $\pm$0.04 & 0.09 	 \\
Ba\,{\sc ii} &  $\pm$0.01   &     $\pm$0.06 &   $\pm$0.25	&  $\pm$0.03     & 0.26 	 \\
La\,{\sc ii}  &  $\pm$0.01   &     $\pm$0.08 &   $\pm$0.03	&  $\pm$0.03 & 0.09 	 \\
Nd\,{\sc ii}  &  $\pm$0.01   &     $\pm$0.08 &   $\pm$0.06	&  $\pm$0.04 & 0.11 	 \\
Eu\,{\sc ii}  &   $\pm$0.01 &     $\pm$0.07 &   $\pm$0.02    &  $\pm$0.03  & 0.08   \\
\\ \hline 

\end{tabular}
\end{table}

% \onecolumn

\begin{table*}
\caption{\label{kine_sample}Kinematical parameters for the sample stars.}
\begin{center}    
\begin{tabular}{llccccccc}

\hline \\
 Star  & v$_{\rm rad}$ (km/s)  & v$_{\rm rad}$ (km/s)  & U   &  V  & W  & P$_{\rm thin}$ & P$_{\rm thick}$ & P$_{\rm halo}$  \\
                & Gaia DR3         & (spec)   &   (km/s)              &     (km/s)       &     (km/s)    & (\%)     & (\%)      & (\%)       \\
 \\ \hline \\

 \multicolumn{9}{c}{{Thin disk stars}} \\
  V354 And       &-12.83$\pm$3.68  &  -7.6    &   -5.5 $\pm$     0.5 	 &     -15.3 $\pm$     0.8	 &       -5.6  $\pm$    0.4  &   99.2 &	     0.8 &	     0.0   \\
 AY Ari         & 22.01$\pm$0.13   & 21.5    &  -55.0  $\pm$    0.8   	 &     -51.2 $\pm$     0.5	 &       -4.4 $\pm$     0.4  &    96.5 &	     3.4 &	     0.1  \\
 EU Dra         & -62.51$\pm$6.47$^a$  &  -57.2   &    28.1 $\pm$     0.2	 &     -49.0 $\pm$     0.7	 &      -28.4  $\pm$    0.7  &    94.7 &	     5.3 &	     0.0   \\
 VW Dra         & 14.43$\pm$0.12    &  14.1   &   -11.2  $\pm$    0.0 	 &      -2.7 $\pm$     0.8	 &       28.6  $\pm$    0.6  &    95.6 &	     4.4 &	     0.0   \\
 V894 Her       & -36.60$\pm$1.09  &   -40.3  &   -18.1 $\pm$     0.6	 &     -18.6 $\pm$     0.5	 &      -31.8 $\pm$     0.7  &    98.0 &	     2.0 &	     0.0  \\
 V991 Her       & -33.42$\pm$0.19    & -33.1    &    10.6 $\pm$     0.7	 &     -64.5 $\pm$     0.7	 &       16.4  $\pm$    0.2  &    88.8 &	    11.1 &	     0.1   \\
 BN Lyn         & ---              &  24.4   &  -32.0  $\pm$    0.9  	 &     -50.8  $\pm$    1.2	 &       -0.4 $\pm$     0.7  &    97.2 &	     2.8 &	     0.0  \\
 CD Psc         & 1.84$\pm$10.31   & -5.1     &   -36.9 $\pm$     0.6	 &      -8.7 $\pm$     0.5	 &        9.8  $\pm$    0.8  &    98.7 &	     1.3 &	     0.0   \\
 EZ UMa         & 16.38$\pm$1.18   & 9.8      &    -26.0 $\pm$     0.7	 &     -33.9 $\pm$     0.6	 &        8.5 $\pm$     0.7  &    98.1 &	     1.9 &	     0.0  \\ 
 RX Vir         & 11.37$\pm$0.19   & 11.0    &   -30.6 $\pm$     0.2    &     -16.3  $\pm$    0.5	 &        7.0 $\pm$     0.8  &    98.8 &	     1.2 &	     0.0  \\ \\
\multicolumn{9}{c}{{Thick disk star}} \\
 AC Aqr         & -5.37$\pm$0.77   &  -9.2    &    122.8  $\pm$   10.7	 &    -109.1 $\pm$     8.6	 &        9.0  $\pm$    0.7  &    2.1 &	    93.5 &	     4.4  \\ \\
\multicolumn{9}{c}{{Halo stars}} \\
  LS Aqr         & -311.56$\pm$0.67  &   -310.3 &   -22.1 $\pm$     4.2	 &    -255.0 $\pm$    11.3	 &      228.7  $\pm$    4.1  &    0.0 &	     0.0 &	   100.0   \\
 HP Eri         & 19.89$\pm$0.25   &   18.4   &   168.2  $\pm$    3.7	 &    -137.7 $\pm$     2.6	 &     -145.5  $\pm$    2.8  &     0.0 &	     4.0 &	    96.0   \\
 KR Vir         & 140.41$\pm$0.31  &   140.4  &   273.0 $\pm$    10.8	 &    -273.0 $\pm$     9.6	 &      -30.0  $\pm$    6.9  &    0.0 &	     0.0 &	    100.0   \\ \\
\hline

\end{tabular} \\
\end{center} 
\footnotesize{a : Radial velocity of EU Dra is from Gaia DR2}\\
\end{table*}

% \twocolumn

\section{Previously analysed SRd variables}

Our interest in  previously analysed  SRd-SRd: variables is (i) to compare results for stars in common with our sample in which previous authors adopted similar spectra and analytical techniques to us. 
 Previous studies  include those by \cite{Andri85, Andri2007}, \cite{Giridhar98, Giridhar99, Giridhar2000},  \cite{Brita2010, Brita2012} and (ii) to extend the sample of variables with a known chemical composition. 
Several stars in  previous studies -- FT Cnc, XY Aqr, RX Cep and Z Aur -- have now lost their SRd status in that they 
are not listed in the {\it GCVS} 2022 catalogue as SRd (or SRd:) variables. Such a revision of status is not surprising given the small amplitude of some photometric  variations. In addition, `variable' lacks a strict quantitative definition and, thus, is most difficult to apply to giant stars; all of which may be subject at a minimum to slight photometric variations of asteroseismic origin and, thus, are almost certainly variable when examined with precision.

Three of our variables were analysed previously from optical spectra: VW Dra \citep{Brita2012}, V894 Her \citep{Brita2010} and RX Vir \citep{Andri2007}. (RX Vir (see Table \ref{atm_warmSRD}) is better linked to the main sequence than to the red giant branch but is a useful check on analytical procedures). The 2012 results and ours for VW Dra are in fine accord as to atmospheric parameters and to abundances [M/H]: the mean [M/H] difference between us and \cite{Brita2012} is $\Delta$ = $+$0.01$\pm$0.14 dex for 21 entries. Extremes contributing to this pleasingly small mean are $-0.26$ from Co\,{\sc i} and $+0.37$ from Nd\,{\sc ii}. (For four light elements, \cite{Brita2012} applied non-LTE corrections.)  Similar remarks  apply to the comparison between our results and \cite{Brita2010}: the mean $\Delta$[M/H] = $-0.09\pm0.21$ from 21 species with extremes $-0.50$ for Ti\,{\sc ii} and $+0.37$ for Nd\,{\sc ii}. With respect to [M/Fe], the mean difference would be increased to near zero by recognizing the overall difference in [Fe/H] between the analyses. 
Finally, the mean difference between our results and those of \cite{Andri2007} for RX Vir is $\Delta$[M/H]  = $+$0.00$\pm$011 dex for 18 entries.

It is of interest to examine recent large surveys of red giants, {\it LAMOST} and {\it APOGEE} \citep{Majewski2017,Wilson2019,Gunn2006, Holtzman2010, Bowen1973, Zasowski2017, Beaton2021, Santana2021, Abdurro2022} for additional  SRd variables.  Given the modest differences in [M/Fe]  between disk and halo populations, we consider entries in the medium-resolution {\it LAMOST} catalogue but not the low-resolution {\it LAMOST} catalogues: then, the only SRds with reported {\it LAMOST} abundances are W LMi and HX Lyn. W LMi is the Li-rich variable on the AGB analysed previously by \cite{Giridhar2000}. HX Lyn has not been previously analysed. For W LMi, the [Fe/H] and four [M/Fe] entries in common between the two studies are in satisfactory agreement. Our evaluation of the Galactic kinematics places HX Lyn with the thin disk and W LMi with the halo.

 The most recent or  DR17 {\it APOGEE} catalogue\footnote{https://data.sdss.org/sas/dr17/apogee/spectro/aspcap/dr17/ \\ synspec\_lte/allStar\-dr17\-synspec\_lte.fits} used here is based on the high-resolution H-band infrared spectra, the stellar parameters and chemical abundances derived  using the Synspec (LTE) spectral synthesis code includes compositions of four SRd or SRd: variables among them AY Ari and RX Vir in Table \ref{warm_srd_table}.  The mean [M/H] difference for AY Ari between us and  {\it APOGEE} for [M/H]  is $\Delta$ = $-$0.03$\pm$0.11 dex for 13 entries. Similarly, the mean difference for RX Vir, the near-main sequence star, is $\Delta$ = $-$0.07$\pm$0.11 dex for 11 entries. (An {\it APOGEE} catalogue is also provided with NLTE abundances  for Na, Mg, K and Ca. NLTE corrections are estimated to be small.) 
These comparisons suggest that our abundance analysis and those provided by the {\it APOGEE} survey \citep{Garcia2016, Shetrone2015,Smith2021, Jonsson2020} are on the same scale.  The {\it APOGEE} catalogue provides compositions for  additional semi-regular variables with {\it GCVS} certification: QU Boo (thin disk) and V335 UMa (thin disk) with their Galactic population indicated.

Thus, comparisons of abundance results for common stars suggest that our  results from optical spectra and published results from {\it APOGEE} spectra are on very similar scales.  This result encourages confidence that abundances for semi-regular variables  -- certainly, those derived from optical spectra free of TiO molecular lines -- may be put up against trends for abundance trends within each of the Galactic kinematical populations  to look for   abnormalities possibly related to   the cause and execution of semi-regularity. Table \ref{previous_srd} lists the previously observed and analysed variables chosen to enlarge the sample of semi-regular variables.

\begin{table}
\caption{\label{previous_srd}Previously analysed SRd stars.}
\begin{tabular}{llcc}
\hline \\
Star  &  Type  &  [Fe/H]  & Reference  \\
\\ \hline \\

 \multicolumn{4}{c}{{Thin disk stars}} \\
QU Boo &	  SRd &	  -0.33  &	APOGEE  \\
V335 UMa &	 SRd: &	  -0.08  &	APOGEE	  \\ 
CW CVn &	 SRd &		-0.14 &	Bri10$^{a}$   \\
V463 Her &	 SRd &		0.06  & Bri10  \\
MS Hya &	 SRd &		-0.33  & Bri10  \\
MQ Hya  &	 SRd  &	  -0.12  & Bri12$^{b}$    \\
VV LMi &	  SRd &		-0.14 & Bri12  \\
HX Lyn &	  SRd: &   -0.483  & LAMOST  \\ \\

\multicolumn{4}{c}{{Thick disk star}} \\
WW Tau &	  SRd &	   -1.1  &	   Gir00$^{c}$  \\ \\

\multicolumn{4}{c}{{Halo stars}} \\
AB Leo &	  SRd  &    -1.96 &	   And07$^{d}$  \\
SV UMa &	  SRd &		-1.87 &	   And07  \\
TY Vir &	  SRd &		-1.71  &   And07  \\
CK Vir &	  SRd &		-1.88  &   And07  \\
WY And &	  SRd &		-1.0  &	   Gir99$^{e}$  \\
VW Eri  &	  SRd &		-1.8  &	   Gir99  \\
UW Lib &	  SRd &		-1.3  &	   Gir99  \\
KK Aql &	  SRd &		-1.2  &	   Gir00  \\
AG Aur &	  SRd &		-1.7  &	   Gir00  \\
W LMi &	      SRd &		-1.1  &	   Gir00  \\

\hline
\end{tabular} \\
\footnotesize{
a : \citep{Brita2010}, b : \citep{Brita2012} \\
c :\citep{Giridhar2000}, d : \citep{Andri2007} \\
e :\citep{Giridhar99} } \\
\end{table}

\section{Composition of Galactic halo, thick and thin disk}

 The  semi-regular variable red giants  span a range in metallicity [Fe/H] across the Galactic halo and the thin and thick disks with  relative abundances   [M/Fe] differing in    ways across the Galactic populations,  as previously determined from  abundance studies based  on  high-resolution optical spectroscopy of F and G dwarfs, for example, by \cite{Reddy2006} (and references therein) and of red giants, as, for example, collated by Nikos Prantzos and illustrated in the review by \cite{Arcones2023}.  Our  goal here is  to test the idea that,  warm semi-regular variable red  giant stars may share the abundance pattern [M/Fe] of  normal red giant stars of the same Galactic population, i.e., the origin of the star's semi-regularity appears not to have induced any anomalies in the abundances of observable elements Na and heavier.  Our principal source of compositions for standard giant stars is the  {\it APOGEE} catalogue which has been itself the source of investigations of the compositions of Galactic populations (for example, \cite{Imig2023}; \cite{Weinberg2022}).  The {\it APOGEE} catalogue is based on  high-resolution H-band infrared spectra. Data from optical data might have served as a reference for the semi-variable stars. Collections of  abundance data for giants and dwarfs in the solar neighbourhood drawn from optical spectra  have been compared with abundance data from  {\it APOGEE}.  \cite{Jonsson2018} compare  extensively several optical collections of abundance data with {\it APOGEE} data from data releases DR13 and DR14 and show that {\it APOGEE} - Optical abundance systematics for our elements-- Na, Mg, Al, Si, Ca, Cr, Mn and Ni -- are less than 0.05 dex (median) and random differences are less than 0.15 dex (std. deviation).

The abundance pattern for giant stars  drawn by us from {\it APOGEE} was based on a few selection rules: a high quality H-band spectrum selected making use of various flags -- EXTRATARG flag, STARFLAG flag, SN\_BAD bit and STAR\_BAD bit of ASPCAP flag and also the bitmasks associated with the elemental abundances --  provided in the catalogue, surface gravity $\log$ g $\leq 3.0$, stars within 1 kpc of Sun with an error smaller than 10 \% in their parallax. Galactic kinematics UVW were computed from Gaia DR3 entries and assignment to a particular Galactic population (halo, thick and thin disk) was assigned if the membership were calculated to be 90\% or greater.  This exercise resulted in a sample containing almost 4860  thin disk, just under 200 thick disk and almost two dozen  halo giants. The entire sample (SRd variables and the APOGEE giants) spans the [Fe/H] range from about $+0.4$ to almost $-2.0$ with the exception of the SRd AC Aqr at [Fe/H] $= -2.3$, which appears to be an  odd thick disk representative.

Compositions for the different Galactic populations were examined in the [M/Fe] vs [Fe/H] plots.  Our principal aim was to assess if the [M/Fe] measures for the semi-regular variables follow the pattern of results for  the {\it APOGEE} giants. {\it APOGEE} due to lack of absorption lines for heavy elements sampling  $s$- and $r$-process elements is unavailable as an adequate reference for heavy elements.  Hence, for heavy elements we drew on  published  abundances from optical spectra.  Our intent here  is {\it not} to interpret in the astrophysical sense the [M/Fe] vs [Fe/H] relations.  Our primary aim is to address the question -- do the variables in terms of their [M/Fe] vs [Fe/H] relations follow the relations exhibited by {\it APOGEE} giants spanning the same [Fe/H] range?

Our selection of semi-regular variables does not extend to the extreme metal-poor halo giants. Yet, a potential indicator of an  unusual nature for semi-regular variables would be a hint that slope of one or more of present trends of [M/Fe] with [Fe/H] suggest a [M/Fe] mismatch with values obtained among extreme metal-poor halo giants. As  representative of analyses of these halo giants, we consider  \cite{Barklem2005}'s  survey of abundance ratios for metal-poor halo stars which begins at about [Fe/H] $\simeq -1.5$ and extends down to $-3.5$. For  the many common elements between our two analyses,  \cite{Barklem2005}'s [M/Fe] at [Fe/H] $\simeq -2$ merge smoothly with the trends shown by semi-regular variable giants.

\begin{figure*}
    \centering
     \includegraphics[width=14 cm, height=20 cm]{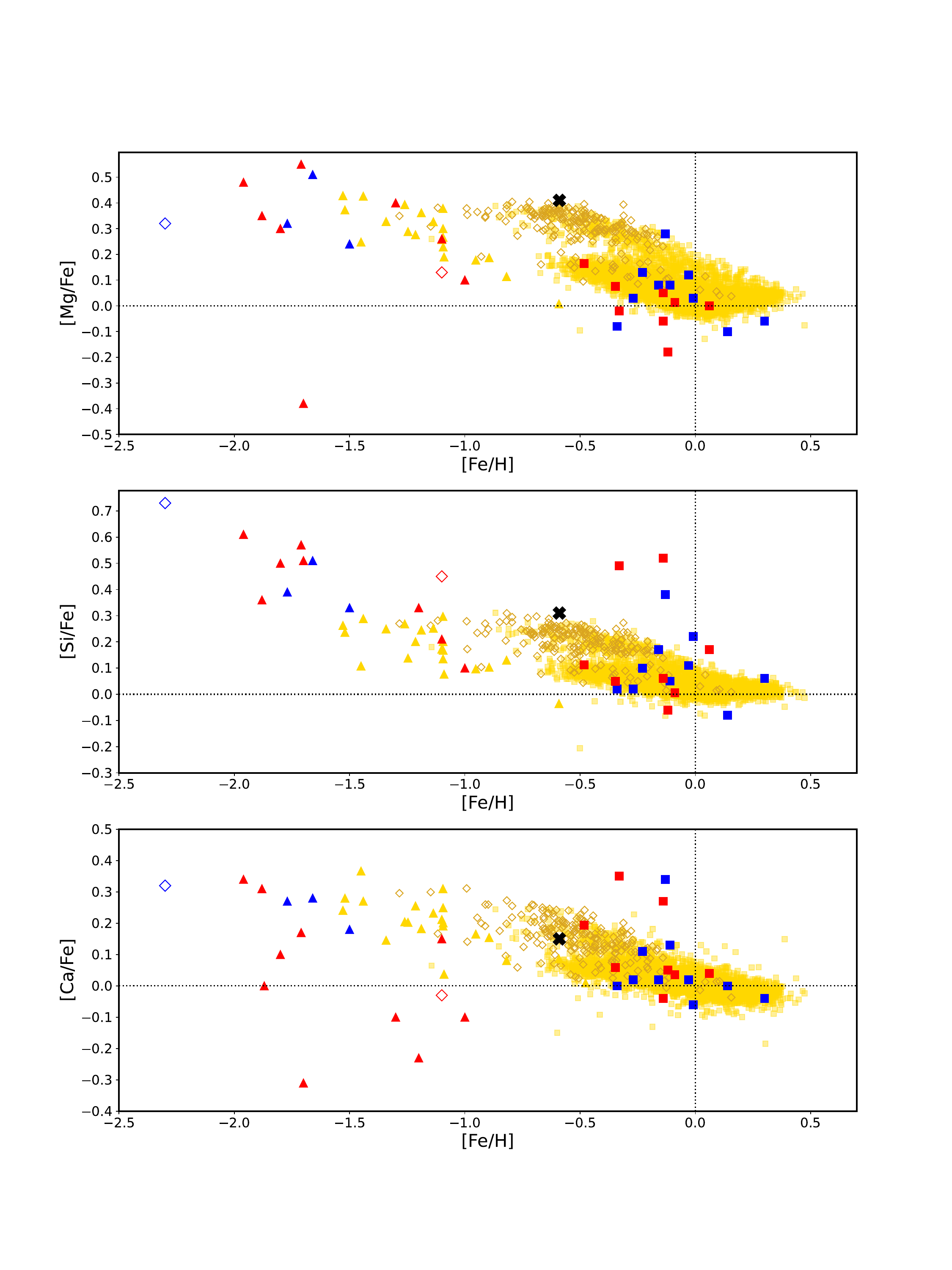}
     %\includepdf[width=15 cm, height=20 cm]{Fig1MgSiCa.pdf}

     \caption{Plots of the abundance ratio [M/Fe] of Mg, Si and Ca against [Fe/H] are shown for the three Galactic populations: thin disk by filled squares, thick disk by open diamonds and halo by filled triangles. Our current SRd variables are represented in blue, previously published SRd variables in red and {\it APOGEE} giants are coloured yellow. Arcturus is represented by the black X.}
    \label{abu_fig1}
\end{figure*}

\begin{figure*}
    \centering
     \includegraphics[width=17 cm, height=17 cm]{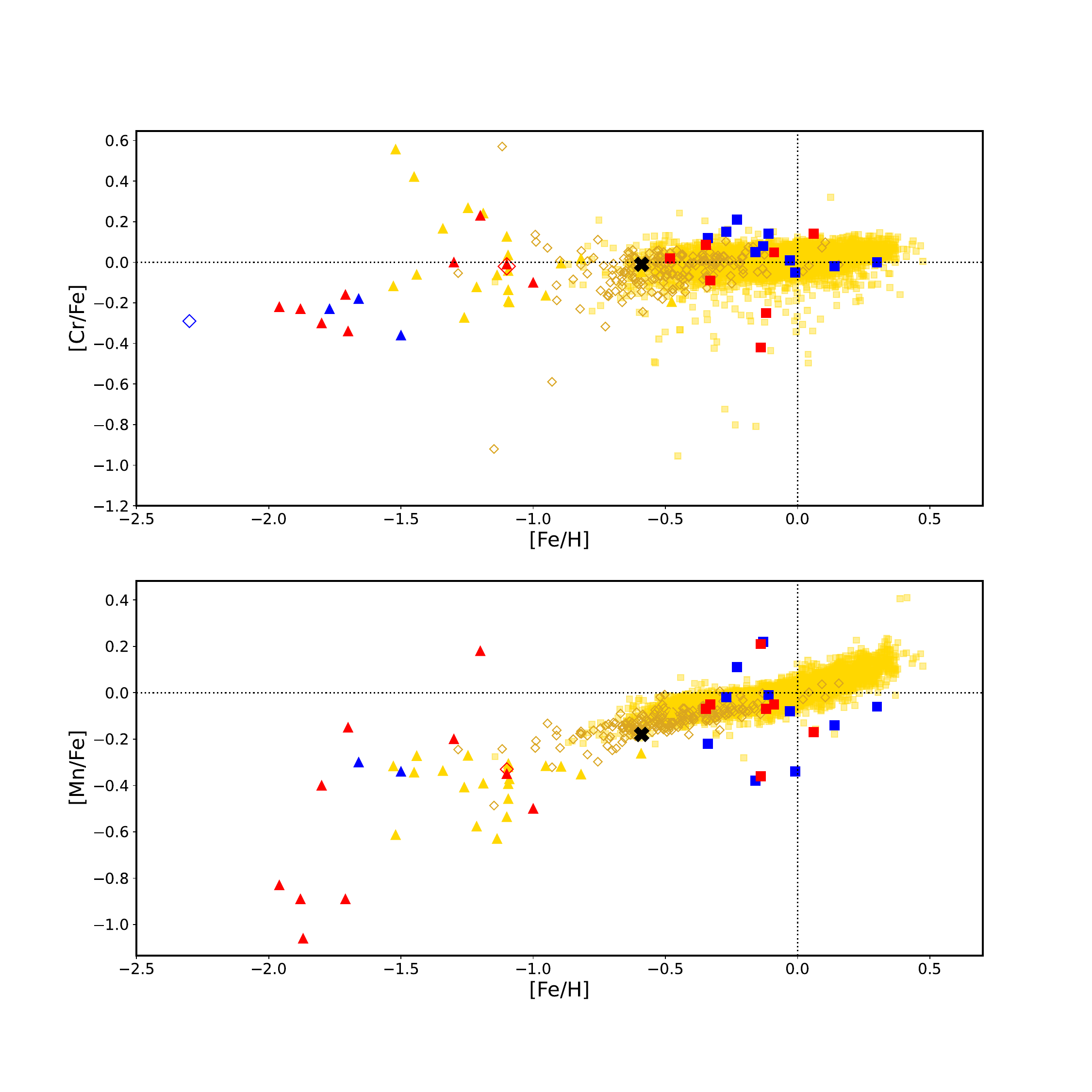}
     %\includepdf[width=15 cm, height=15 cm]{Fig2CrMn.pdf}
     \caption{Abundance ratio [M/Fe] of Cr and Mn against [Fe/H]. Symbols and colour codes are the same as in the previous figure.}
    \label{abu_fig2}
\end{figure*}

Although the sample Mg to Ni was examined individually, [M/Fe] vs [Fe/H] plots (Fig. \ref{abu_fig1} and Fig. \ref{abu_fig2})  are shown only for M = Mg, Si, and Ca in Fig. \ref{abu_fig1} and M = Cr and Mn in Fig. \ref{abu_fig2}.  In these Figures, the thin disk giants selected from {\it APOGEE} DR17 (Synspec LTE) catalog are represented by yellow filled squares, the thick disk by yellow open diamonds and the halo by yellow filled triangles (A possibly common misapprehension is to identify the thin disk  with the low [Mg/H] vs [Fe/H] strip and the high [Mg/H] vs [Fe/H] strip almost exclusively with the thick disk. But Fig. \ref{abu_fig1} following, for example,  \cite{Imig2023} and \cite{Weinberg2022} shows the composition and kinematics of Galactic disk to be  complex and, in particular, the composition of the thin disk appears to be binary in nature). Superimposed on each panel in these figures are results for the semi-regular variables: blue and red symbols distinguish our variables from those previously analysed, as for the {\it APOGEE} giants,   variables from the thin disk are represented by filled squares, thick disk members by open diamonds and halo representatives by filled triangles.

Fig. \ref{abu_fig1}'s value is that normal red giants as observed in {\it APOGEE} and earlier reported by many surveys show  similar [M/Fe] vs [Fe/H] relations for M = Mg, Si and Ca across the three Galactic populations with small scatter in the [M/Fe] at a given [Fe/H] in the halo. This common behaviour across the [M/Fe] vs [Fe/H] relations eases an identification of unusual results for the semi-variables. For M = Mg,  semi-variables with the sole exception of the odd-appearing halo star AG Aur ([Mg/Fe] =  -0.4 at [Fe/H] $= -1.8$ from \cite{Giridhar2000}) match satisfactorily the behaviour of the {\it APOGEE} halo giants. The optical spectrum of  AG Aur was, as noted above, marred by TiO bands which may account for the low Mg abundance. The [Si/Fe] panel shows a similarly tight distribution but for our thick disk giant AC Aqr with [Si/Fe] $=+0.73$ and three thin disk SRd variables (two from previous studies and our V991 Her) of above average [Si/Fe]. Ca shows a more severe introduction of unusually low [Ca/Fe] for several  [Fe/H]-poor  SRd variables  including not only AG Aur. The other  aberrant [Ca/Fe] indices are  from spectra with prominent TiO bands;  presence of TiO bands in spectra of cooler semi-variables may result in additional unappreciated spectral complexity in optical spectra  affecting line identification and measurement. 
Three thin disk variables exhibit   [Si/Fe] and [Ca/Fe] of  greater than normal indices but not apparently [Mg/Fe]:  our V991 Her and  the previously analysed MS Hya (\citep{Brita2010}) and VV LMi (\citep{Brita2012}.

Our second illustration involves Cr and Mn from the iron group (Fig. \ref{abu_fig2}). Both elements show a weak trend of [M/Fe] with [Fe/H].   [Cr/Fe] across the semi-regular variables which overlaps   the weak tight trend displayed by the  {\it APOGEE} giants. A few {\it APOGEE} giants have apparently unusually either high or low [Cr/Fe]. The sole previously observed SRd with even a mildly  low [Cr/Fe] is  VV LMi at [Fe/H] $= -0.1$ with [Cr/Fe] $= -0.4$ with its Cr abundance based on 21 Cr\,{\sc i} lines \citep{Brita2012}). Indeed, isolation of Cr in these plots is consistent with the assessment of the `reliability' of {\it APOGEE} abundances\footnote{https://www.sdss4.org/dr17/irspec/abundances/}: Cr is identified as `less reliable', Ca as 'reliable', and Mg, Si, Mn  and Fe as `most reliable'.

Perhaps,  the most puzzling example of an anomaly in the entire [M/Fe] vs [Fe/H] plane is provided by the four previously-observed halo variables at [Fe/H] $\sim -1.9$ with their remarkable  [Mn/Fe] about 0.7 dex less than the trend set by halo giants and  halo variables at the same [Fe/H]. This quartet were provided by \cite{Andri2007} who drew attention to them: "All [four] SRd program stars show uniform [Mn/Fe] ratios that are typical for the stars with metallicity [Fe/H] $\simeq -3.5$. The reason for such a low Mn abundance in these stars is unknown." For SV UMa, one of four variables, \cite{Giridhar98} had reported a `near-normal' [Mn/Fe] $= -0.2$ at [Fe/H] $= -1.4$ in disagreement with Andrievsky et al. as to Mn and Fe abundance.  This Mn puzzle remains and warrants further spectroscopic examination. \cite{Barklem2005}, as \cite{Andri2007} noted,  provide a map of [Mn/Fe] in red giants to the  Fe limit of the halo ([Fe/H] $\simeq -3.5$)  which merges smoothly with the  data in Fig. \ref{abu_fig2} less the troublesome quartet.  The high [Mn/Fe] $= -0.2$ for the halo  semi-variable KK Aql \citep{Giridhar2000} also warrants additional investigation.

 Agreement between compositions of semi-variable giants and regular giants (and dwarfs) extends to elements beyond the Fe-group. {\it APOGEE}'s selection of heavy elements is restricted to Ce so our comparisons are based on optical spectra.  For thin disk semi-variable giants, our sample of seven provide a mean [Eu/Fe] $= 0.12\pm0.06$ in satisfactory agreement with the result from previously analysed variables giving [Eu/Fe] $= -0.08\pm0.12$ and in fine agreement with normal giants and F-G dwarfs (e.g., \cite{Reddy2006}) over the range [Fe/H] $\ge -0.5$. (CW CVn with [Eu/Fe] $= 0.64$ \citep{Brita2010} appears to be an anomaly).  This agreement over [Eu/Fe] estimates extends down close to [Fe/H] $\sim -1.0$ with the possible exception of W LMi at [Fe/H] $= -1.1$ where [Eu/Fe] $= +0.9$ \citep{Giridhar2000} for this super Li-rich early-AGB variable. Eu is the  classical indicator of an {\it r}-process contribution: thus, the {\it r}-process contribution at a given [Fe/H] to normal dwarfs, giants and the semi-variable giants appears similar for the halo and the disks.  For the  measured heavy elements with an anticipated major origin from the $s$-process -- Y, Zr, Ba and  La with about 75\% and Nd with 50\% {\it s}-process to their solar abundance -- the [M/Fe] across the semi-variables and regular dwarfs and red giants is approximately flat but rising slightly at the lowest [Fe/H] as the expected {\it r}-process contribution increases -- say, from Y and Zr to Nd and Eu.  Again, similar [M/Fe] for semi-regular variables and `normal' giants appears likely.

Inspection of Fig. \ref{abu_fig1} and Fig. \ref{abu_fig2} suggests that the compositions of semi-regular variables -- analysed here and previously reported -- closely resembles those of red giants as reported by {\it APOGEE}; both groups of giants share  a common origin. Exceptions found among individual  [M/Fe] vs [Fe/H] trends from both optically analysed and {\it APOGEE} spectra deserve further  spectroscopic observations and theoretical considerations.  They highlight that stellar compositions remain a challenging pursuit.

\section{Concluding remarks}

Semi-regular red giant variables such as those in Table \ref{warm_srd_table} likely owe their variability to non-radial pulsations of primarily long-period and low amplitude. Variability is thus a natural phenomenon expected of all giants with a photometric amplitude increasing with decreasing effective temperature, that is from G to K to M type (e.g., \cite{Jorissen97}; \cite{Henry2000}).  Detailed observational photometric and/or radial velocity variability studies of these red giants provide an interior probe  allowing a distinction to be made between a star evolving up the red giant branch  and one post He-core flash and now on the asymptotic giant branch.

Our goal with this abundance analysis of a selection of K giant SRd and SRd: variables was to search for abundance differences among elements of Na and heavier between our sample of  cataloged semi-variables and similar K giants thought to be photometrically constant. \cite{Henry2000} report that many K giants are photometrically constant at  about the 0.002 mag level. To the precision of the spectroscopic abundance analyses -- optical and infrared -- the elemental abundances of semi-variable giants drawn from  the Galactic halo, thick and thin disks show no systematic differences with the (apparently) photometrically constant red giants from the same Galactic population.

It remains to be seen if these seemingly photometrically contrasting groups of K giants show abundance differences for elements lighter than Na, as, for example, Li, C, N, O and F and their isotopes. Some differences are anticipated, if only, because the sample of giants may include stars from both red giant and asymptotic giant branch and the associated evolutionary structural changes (e.g., the He-core flash marking the birth of the low-mass asymptotic giant branch giant) may result in light element and isotopic  abundance changes within a giant's atmosphere.  Thorough  spectroscopic investigation of  M-type giants  near the tip of the red giant branch and along the asymptotic giant branch calls for high-resolution infrared spectra as provided by {\it APOGEE} and {\it IGRINS} \citep{Park2014, Yuk2010} but the intriguing element Li, which eludes the infrared observer, demands an optical spectrum! \\

\onecolumn

\clearpage

\twocolumn

{\bf DATA AVAILABILITY} \\
The Survey data used in this article can be accessed at the following links. \\
SDSS DR17 for APOGEE catalog :\\
\url{https://data.sdss.org/sas/dr17/apogee/spectro/
aspcap/dr17/synspec\_lte/allStar-dr17-synspec\_lte.fits}  \\
LAMOST DR8 Medium Resolution Catalog :\\
\url{http://www.lamost.org/dr8/}    \\
GAIA DR3 catalog :\\
\url{https://gea.esac.esa.int/archive/} \\ \\

{\bf ACKNOWLEDGEMENTS} \\

We thank our friend N. Kameswara Rao for his suggestion that SRd variables were  worthy of pursuit with the {\it Robert G. Tull} spectrograph. We note too with much appreciation the insights into the voluminous records of the {\it APOGEE } project volunteered by  Carlos Allende Prieto and Verne V Smith. We thank the referee for a thoughtful review.

This work has made use of data from the European Space Agency (ESA) mission
{\it Gaia} (\url{https://www.cosmos.esa.int/gaia}), processed by the {\it Gaia}
Data Processing and Analysis Consortium (DPAC,
\url{https://www.cosmos.esa.int/web/gaia/dpac/consortium}). Funding for the DPAC
has been provided by national institutions, in particular the institutions
participating in the {\it Gaia} Multilateral Agreement.

Guoshoujing Telescope (the Large Sky Area Multi-Object Fiber Spectroscopic Telescope LAMOST) is a National Major Scientific Project built by the Chinese Academy of Sciences. Funding for the project has been provided by the National Development and Reform Commission. LAMOST is operated and managed by the National Astronomical Observatories, Chinese Academy of Sciences.

Funding for the Sloan Digital Sky 
Survey IV has been provided by the 
Alfred P. Sloan Foundation, the U.S. 
Department of Energy Office of 
Science, and the Participating 
Institutions. 

SDSS-IV acknowledges support and 
resources from the Center for High 
Performance Computing  at the 
University of Utah. The SDSS 
website is www.sdss4.org.

SDSS-IV is managed by the 
Astrophysical Research Consortium 
for the Participating Institutions 
of the SDSS Collaboration including 
the Brazilian Participation Group, 
the Carnegie Institution for Science, 
Carnegie Mellon University, Center for 
Astrophysics | Harvard \& 
Smithsonian, the Chilean Participation 
Group, the French Participation Group, 
Instituto de Astrof\'isica de 
Canarias, The Johns Hopkins 
University, Kavli Institute for the 
Physics and Mathematics of the 
Universe (IPMU) / University of 
Tokyo, the Korean Participation Group, 
Lawrence Berkeley National Laboratory, 
Leibniz Institut f\"ur Astrophysik 
Potsdam (AIP),  Max-Planck-Institut 
f\"ur Astronomie (MPIA Heidelberg), 
Max-Planck-Institut f\"ur 
Astrophysik (MPA Garching), 
Max-Planck-Institut f\"ur 
Extraterrestrische Physik (MPE), 
National Astronomical Observatories of 
China, New Mexico State University, 
New York University, University of 
Notre Dame, Observat\'ario 
Nacional / MCTI, The Ohio State 
University, Pennsylvania State 
University, Shanghai 
Astronomical Observatory, United 
Kingdom Participation Group, 
Universidad Nacional Aut\'onoma 
de M\'exico, University of Arizona, 
University of Colorado Boulder, 
University of Oxford, University of 
Portsmouth, University of Utah, 
University of Virginia, University 
of Washington, University of 
Wisconsin, Vanderbilt University, 
and Yale University.

%\bibliographystyle{mn2e}
%\bibliography{SRD_ref} 

\label{lastpage}

\end{document}